# Reaction-Diffusion Driven Patterns in Immiscible Alloy Thin Films

Vivek C. Peddiraju, Shourya Dutta-Gupta, Subhradeep Chatterjee*
Materials Science and Metallurgical Engineering, Indian Institute of Technology Hyderabad, Sangareddy 502284, Telangana, India
*Corresponding author e-mail: subhradeep@msme.iith.ac.in

## Abstract

Controlling the microstructure of thin films is of critical importance for various applications. We demonstrate a methodology for tuning the local microstructure through film-substrate interactions using Ag-Cu as a model system. Metastable single-phase Ag-Cu thin films are deposited on Si substrates pre-patterned by FIB milling. During post-deposition annealing, localized film-substrate reaction around the milled patterns produces a distinct microstructure termed as the 'halo'. It consists of copper silicide and almost pure Ag, while the far-field film forms a random mixture of Cu and Ag-rich domains through phase separation. We show that the extent of the halo can be controlled by varying the temperature and duration of annealing. We present a semi-analytical kinetic model of product and halo growth that incorporates species balance, diffusional transport and a modified Stefan condition. Predictions from the model reveal two distinct growth regimes of the product with power law indices of 1/2 and 2/7 and experimental data fall into the latter regime. These regimes originate from the dimensionality of growth (2d or 3d) compared to that of solute transport (2d), which in turn depend on film thickness and species diffusivity. Using an inverse optimization procedure, we also estimate the diffusivity, which suggests grain boundary diffusion to be the dominant transport mechanism. This study provides an avenue and framework for microstructural engineering of alloy thin films through interfacial reaction.

## 1. Introduction

Engineered thin films with tailored microstructures have been extensively used for applications like catalysis, advanced display technologies, semiconducting platforms, sensing and batteries [1–13]. The ability to tune the phase structure, composition and domain size through processing conditions, both *in situ* and *ex situ*, enables property optimization for a given application. Advances in nanofabrication techniques in the last decades have enabled researchers to enhance the control over the properties even further by employing various pattering and structuring techniques at the nanoscale. Physical vapor deposition is one of the popular techniques to produce thin films with precise compositional and thickness control [14–16]. In most cases, the as-deposited films are often thermodynamically metastable due to the rapid immobilization of high energy atoms on to a substrate during the deposition process [15,17,18]. Subsequent thermal treatments provide avenues for the films to relax to a relatively more stable configuration through crystallization, grain growth, phase separation and chemical reactions. Two or more of these processes may take place sequentially or concomitantly depending on their driving forces and kinetic factors. Typically, multiphase films can be produced by exploiting the tendency of the constituents to phase separate, to react chemically with the substrate, or by a combination of both.

Phase separation in films, driven either by nucleation and growth processes or by spinodal decomposition [19], can lead to the formation of dispersed phases in a matrix, layered structures or interconnected domains [20–24]. Parameters such as film thickness, alloy composition, deposition rate, substrate temperature, post-deposition thermal treatment, diffusional mobilities of constituent species, and interfacial energies and elastic moduli of the phases are known to influence the film microstructure [25–28]. For example, Adams *et al.* [29,30] reported a transition from lateral to transverse modulation of the microstructure in Al-Ge thin films with increasing film thickness. Studies on polymeric films have demonstrated that the wavelength of the composition modulation scales with film thickness [31–33]. Capillarity too can influence the separation patterns in a multiphase film. A phase with a lower surface energy tends to form on the free surface and one with a lower interfacial energy with the substrate can preferentially wet the substrate [31,34–39]. These can lead to the formation of layered structures. Capillarity effects can also be induced by pre-patterning of the substrate to alter the wetting behavior [40–43]. The effect of the substrate orientation and epitaxial strain on phase separation in oxide films has also been studied extensively [44–47].

Aside from diverse physical phenomena, chemical reactions can occur between the film and substrate during deposition and/or subsequent annealing treatments. In interconnects used in microelectronics, reaction between metallic films and Si-substrate can lead to the formation of a series of metal silicide layers of different stoichiometries [48–50]. Silicide formation has also been observed in binary alloy thin films on Si. The extent of these reactions can be controlled to produce stratified composite structures consisting of silicide and metallic layers of different chemistry [51–54]. It was shown recently that local microstructural changes in immiscible Ag-Cu thin films could be induced by confining chemical reactions between the film and the Si substrate [55]. The reaction led to the formation of copper silicide surrounded by a silver-rich halo. However, details of the morphological evolution and its dependence on annealing conditions were not reported.

In the present work, we demonstrate that reaction-influenced phase separation can lead to unique patterns in alloy thin films using the Ag-Cu system as a model system. A systematic variation of annealing temperature and time reveals structural and morphological features, as well as quantifies the relationship between annealing parameters and the spatial extent of the reaction-modified zones. Building on experimental observations, we develop a simplified one-dimensional semi-analytical kinetic model of the reaction-diffusion phenomena. It incorporates species conservation and linearized flux balance for an axisymmetric product geometry where growth takes place both laterally and vertically, driven primarily by a lateral species transport through the film. It also revealed two limiting growth regimes with distinct power-law indices. Finally, we used an inverse optimization procedure with the model to obtain solute diffusivity in the film, thus providing a useful estimate of this important material parameter.

# 2. Materials and Methods

### *2.1 Processing*

Elemental Ag and Cu targets (99.99% pure, diameter 50.8 mm, thickness 3 mm) procured from MSE Supplies LLC were used for growing Ag-Cu alloy films. Magnetron co-sputtering of elemental Ag and Cu targets was carried with targets arranged in a confocal geometry; 80 W RF and 45 W DC power were used for Ag and Cu, respectively. The deposition chamber was evacuated to a pressure of $2\times10^{-6}$ mbar before the deposition. A nominal pressure of $4.5\times10^{-3}$ mbar was maintained during the deposition process by purging Ar gas (99.99% pure) into the chamber at a rate of 5 SCCM. Deposition time~~s~~ of 90 s was used to obtain near-equimolar ($Ag_{0.5}Cu_{0.5}$) alloy thin films of nominal thicknesses of $\sim$50 nm.

Si-(100) wafers coated with a 50 nm thick amorphous silicon nitride (ASN) layer obtained from Rouge Valley Microdevices were used as substrates for growing the films. The films were annealed inside the vacuum chamber ($\sim10^{-5}$ mbar) immediately after deposition at temperatures between 150 $^{\circ}$C and 400 $^{\circ}$C.  Note that there will be a slight drop in actual surface temperature of the film compared to the value returned by thermocouple; however, this is estimated to be very small and the same for all the experiments. A uniform heating rate of 10 $^{\circ}$C/min is employed during the ramping up. The films cool down naturally to the room temperature inside the vacuum chamber with an average cooling rate of 10 $^{\circ}$C/min. Preliminary experiments to select time and temperature window for post-deposition annealing were carried out on films grown on bare Si-(100) wafers without passive ASN layer. ASN coated substrates are used for detailed time-temperature study as these provide greater control over localized chemical reaction due to the presence of the $SiN_x$ passivation layer. The lower limits for temperature and time were set by the extent of chemical reaction, whereas the uppers limits were constrained by the onset of dewetting of the films. The chemical reaction sites were pre-patterned using $Ga^+$ focused ion beam (FIB) in  a JEOL JIB-4700F microscope. Multiple circular patterns of diameters 400, 600, 800 and 1000 nm were produced on the substrate using FIB.

### *2.2 Structural Characterization*

The microstructure of the processed films (in both as-deposited and annealed conditions) were studied using electron microscopy. A JEOL JIB-4700F multi-beam scanning electron microscope (SEM) was used for imaging film microstructure. Unless specified, all micrographs presented here were recorded using back scattered electron (BSE) imaging mode. The composition of the films was confirmed using an EDAX Octane Elect Plus energy dispersive X-ray spectrometer (EDS) attached to the SEM. FIB was also used for preparing site-specific transmission electron microscopy (TEM) specimens. The thinned TEM lamella were transferred onto copper half grids using an Oxford Instruments OmniProbe-350 nanomanipulator. These lamellae were analyzed in a JEOL F200 TEM operated at 200 kV using a combination of diffraction and imaging, we well as high angle annular dark field (HAADF) imaging and EDS analysis in the scanning transmission (STEM) mode. Phase analysis of the films was carried out using a Bruker D8 Discover grazing incidence X-ray diffraction (GIXRD) with a Cu-$K_\alpha$ source at a 1° angle of incidence. All patterns were collected within the $2\theta$ window of 30°-90° with 0.05° resolution at a rate of 3 s/step.

### *2.3 Image Analysis*

The microstructure of the phase-separated 'bulk' film consists of Cu-rich domains distributed within an Ag-rich matrix. We carried out additional image analysis for accurately determining the characteristic length scales of the phase-separated microstructures. In order to quantify the changes in the phase separation patterns, we used two separate image analysis methods, *viz.*, image autocorrelation function (ACF) analysis and linear chord length distribution (CLD) analysis for ascertaining the consistency and reliability of the measurements [56]. Further details of these methods are provided in the Supplementary Material along with relevant analyses from simulated test images in Figures S1-S2. For each annealing condition, multiple micrographs were analyzed to determine the size of the compositional domains.

The length scale of the reaction-modified zone (termed as the 'halo') is measured directly from the BSE images using the image analysis software ImageJ. The halo length scale is defined as the distance from the edge of the silicide particle to the halo/bulk film boundary. Since silicide particles have a well-defined geometric shape, an equivalent ellipse is fitted to this particle using the *oval selection* option in ImageJ, and its major and minor axes are used to determine the width of the product phase, as well as the halo length.

## 3. Results and Discussion

Deliberate tuning of microstructural patterns is central to achieving desirable functionalities in a thin film system. In addition to film composition and its processing history, factors like capillarity and interactions with external environment (*e.g.*, through imposed fields or chemical reactions), can be exploited for engineering the film microstructure. For example, multiphase films can be further tailored by selective wetting/dewetting phenomena [57–63] to give rise to desirable microstructures. Chemical reactions can also be utilized to produce new phases and structures. Figure 1 presents a schematic where a metastable homogeneous film phase separates upon heating to produce, say, an overall uniform nanocomposite structure. This uniformity, however, can be altered, by introducing local chemical reactions, which produces new microstructural patterns in the film.

This approach is versatile as different reaction product(s) and reaction-modified structures can be produced by choosing appropriate film-reactant(s) combinations and varying the thermal conditions. Further, the external chemical source can be introduced in predetermined locations in the film through standard lithography techniques. In the present work, we demonstrate the applicability of this fabrication route by choosing Ag-Cu as a model alloy thin film system and Si-substrate as the external reactant.

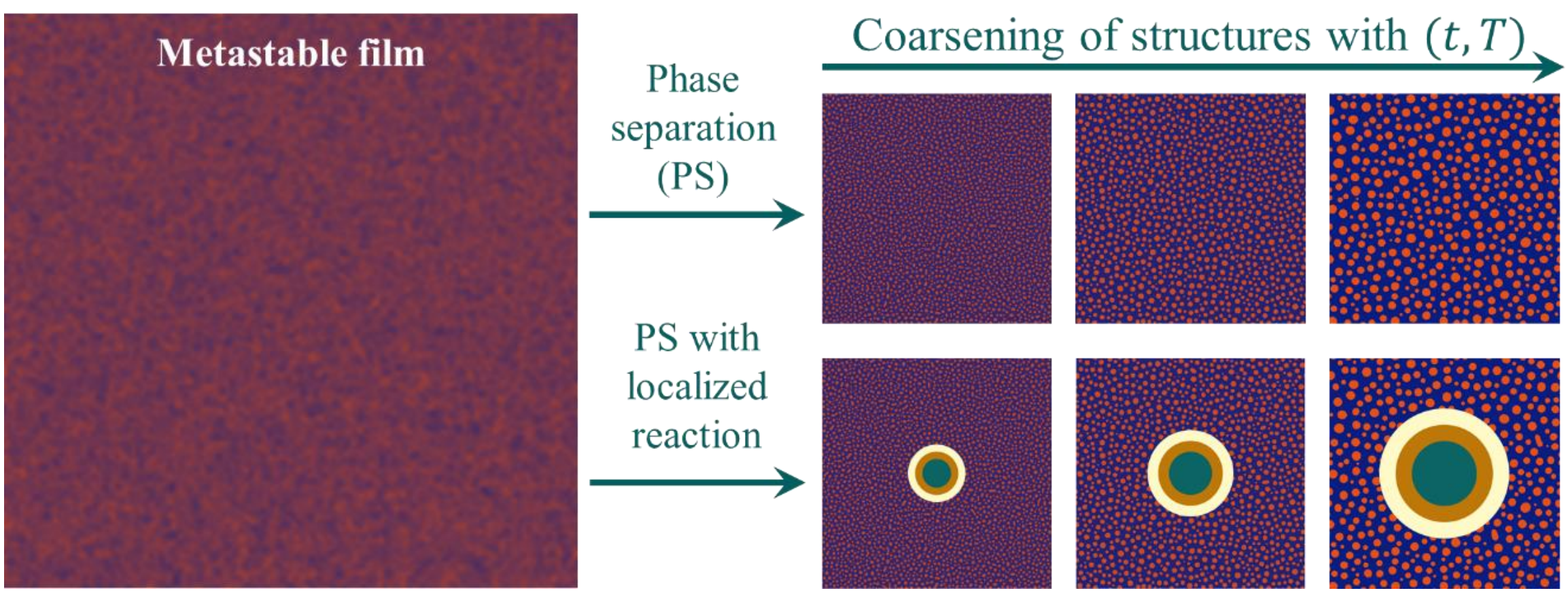

***Figure 1****: Illustration of microstructural engineering of multiphase films through chemical reaction. (Left) As-deposited homogeneous, single-phase film. (Right-top) Evolution of phase separated structure during thermal exposure. (Right-bottom) Local chemical reaction can be introduced to produce a modified microstructural pattern with one or more reaction products.*

### *3.1 Reaction engineering of Ag-Cu thin film*

Figures 2(a-c) shows a schematic of the workflow employed to introduce site-specific chemical reactions between the Ag-Cu film and the Si substrate. FIB pre-patterning was used to locally remove the passive ASN layer and expose the reactive Si-substrate prior to the deposition of the Ag-Cu film. Specific circular patterns (termed as *apertures*) extending slightly into the substrate were used in the current study. The patterning created direct contact between the bare Si and the film where film-substrate reaction could take place during the subsequent annealing treatment.

As shown in Figure 2(d), the as-deposited alloy film has a fine grained microstructure, and further analysis establishes it to be a single-phase metastable FCC solution with a uniform nominal composition of $Ag_{53\pm1}Cu_{47\pm1}$. Figure 2(e) shows a representative micrograph of an annealed film. It confirms that the bulk film decomposes into a mixture Ag- and Cu-rich domains after annealing. It also shows that where the film was in contact with the bare Si (*i.e.*, apertures), chemical reaction between the film and the Si substrate resulted in the formation of copper silicide phase (see **Section 3.2**). The region between silicide and the bulk film undergoes distinctive microstructural changes leading to the formation of a characteristic structure termed as the *halo* [55]. Figure 2(f) shows a circularly averaged radial BSE intensity profile from the center of the silicide of the BSE intensity which reflects the variation of Ag-content as a function of distance. It changes sharply adjacent to the aperture, and fluctuates only slightly about a mean value far away, reflecting the composition variation in the halo and the average composition of the bulk film, respectively. Note that the central silicide appears bright due to the contribution of the topographic contrast (*i.e.*, it projects above the film surface).

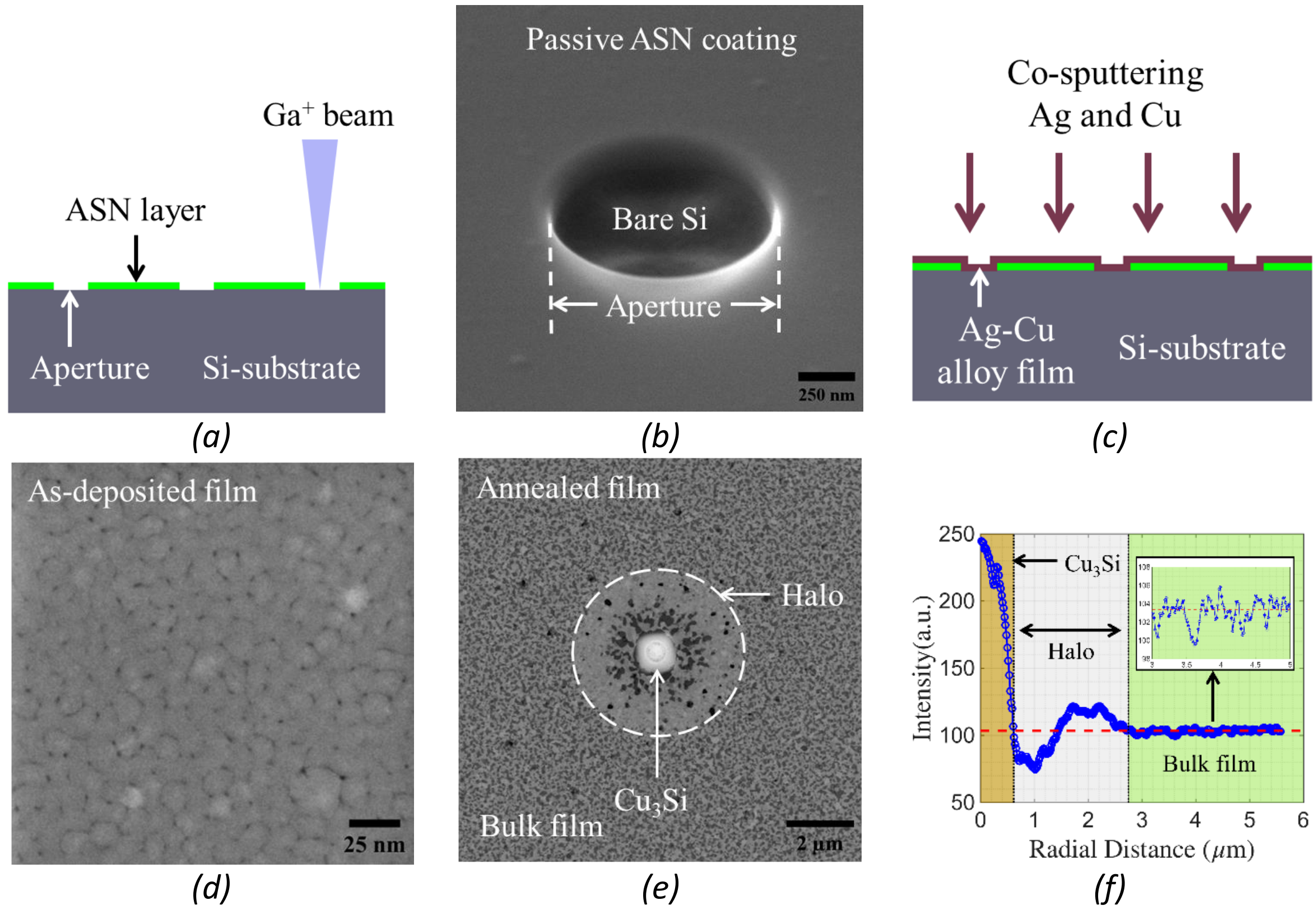

***Figure 2:*** *(a) Schematic of FIB pre-patterning of the substrate. (b) View of a single aperture at 53° tilt; captured using secondary electrons. (c) Schematic of the deposited film on the patterned substrate. (d) STEM image of the film in as-deposited state. (e) SEM-BSE image after annealing showing the formation of copper silicide in the center. Reaction-modified zone marked by the white dashed circle represents the halo made up of an Ag-matrix (bright grey) with secondary coper silicides (dull grey). (f) Circularly averaged intensity profile as a function of radial distance from aperture center demarcates different regions*

### ***3.2 Characterization of reaction product***

The reaction between copper in the film and the Si-substrate results in the formation of a compound in and around the FIB-milled aperture. To determine structure and chemistry of the reaction product formed during annealing, TEM lamella were cut diametrically from the halo region using FIB milling. Figure S3 shows the region from which the TEM specimen is extracted for analysis. Note that the TEM specimen is prepared from a film grown on the bare Si substrate; however, identical morphology and chemistry of the silicide phase were confirmed for both the pre-patterned and bare substrates by SEM and EDS analysis.

Figures 3(a) and 3(b) present the HAADF-STEM image and STEM-EDS maps of the cross-sectional sample. The maps reveal a uniform distribution of Cu and Si inside the particle, suggesting the formation of a copper silicide phase by film-substrate reaction. The composition of phase is found to be $Cu_{79}Si_{21}$, close to a stoichiometry of $Cu_3Si$. Note that the Cu-content is likely to be an overestimate due to the contribution from the Cu half-grid to which the TEM lamella is attached. We also observed the presence of oxygen along the boundary of the copper silicide particle. Furthermore, there is no noticeable Ag inside the particle. We also observed a few tiny particles in regions adjacent to the central copper silicide formed in the Halo region of the structure. The STEM micrograph in Figure 3(c) shows the representative image of these particles. EDS analysis indicates that they have a nominal composition of about $Cu_{25}Si_{75}$. The differences in the relative amounts of Cu and Si in these two silicides are apparent from the EDS profiles shown as inset in Figure 3(c). Since no such Si-rich silicide phase exists in the Cu-Si system, these small particles may be a metastable phase, or what is more likely, the higher Si signal could be from the Si substrate in which they were embedded.

The Cu-Si phase diagram contains six copper silicide phases, *viz.*, $\kappa, \beta, \delta, \gamma, \epsilon$ and $\eta$, with progressively lower Cu:Si ratios [64,65]. The $\eta$ phase with a nominal stoichiometry of $Cu_3Si$ has the lowest Cu:Si ratio and it is the most frequently observed silicide phase to form by solid-state reaction between Cu thin films and Si substrate [66–68]. We use electron diffraction (ED) to validate that the Cu-Si rich particle is indeed $Cu_3Si$. Note that several ED and X-ray diffraction (XRD) studies have been conducted to determine its structure [69–75]. The following points appear to emerge from these studies: (a) $\eta$ belongs to the trigonal crystal system ($P\bar{3}m1$) which can be described by a hexagonal unit cell with a≅4.06 Å and c≅7.33 Å, (b) its structure can be viewed as a stacking of layers along the c-axis, (Figure 3(d) shows projections along $[10\bar{1}0]$ and [0001] directions), (c) the positions of Cu and Si atoms in these layers are not uniquely defined, and (d) there exist low-temperature polymorphs $\eta'$, $\eta''$ and $\eta'''$ whose structures are complex but can be derived from the fundamental hexagonal unit cell representing the trigonal crystal.

Figures 3(e) and 3(f) present the $[11\bar{2}0]$ zone axis ED pattern obtained from the silicide particle shown in Figure 3(a) and the simulated pattern, respectively. Additional ED patterns from other zone axes obtained by systematic tilting to different tilt angles, along with corresponding simulated patterns, sketch of the Kikuchi space swept while collecting the ED patterns and the recorded Kikuchi patterns are provided in Figures S4 and S5. All the fundamental spots in these patterns are verified to originate from the basic hexagonal unit cell. The angles between the zones calculated from recorded tilt angles match with those calculated based on their indices, thus further confirming consistent indexing. The lattice parameters calculated from these patterns are $a$ = 3.96 Å and $c$ = 7.48 Å which differ by ∼2% from the reported values.

In addition to the fundamental spots of $\eta$, additional features appear in the ED patterns. Figure 3(e) shows streaks along the $000l$ reciprocal lattice direction. The centroid of the streaks are located midway between the $000l$ type spots which suggests doubling of the unit cell along the $c$-axis. A similar streaking along the $c$-axis was observed by Wen and Spaepen [70] during $\eta' \rightarrow \eta''$ transformation and they attributed it to the formation of domains with faulted stacking sequence. In the present study, the streaks disappear when the crystal is titled about $c$-axis and brought into the $[10\bar{1}0]$ zone; see Figure S5(b) of SI. Such extinction of streaks in certain zone axes have been reported in diffraction studies on silicides as well as other materials [76–78]. This is usually attributed to the formation of faulted domains with a different stacking sequence. In addition to the streaks, we also observed satellite spots around all fundamental spots along the $1\bar{1}03^*$ direction in the $[11\bar{2}0]$ zone. A similar observation was made by Wen and Spaepen as well for the $\eta' \rightarrow \eta''$ transformation [70]. These satellite spots appear around fundamental spots due to the displacement modulations as reported by Palatinus [72] and Corrêa *et al.* [73,74].

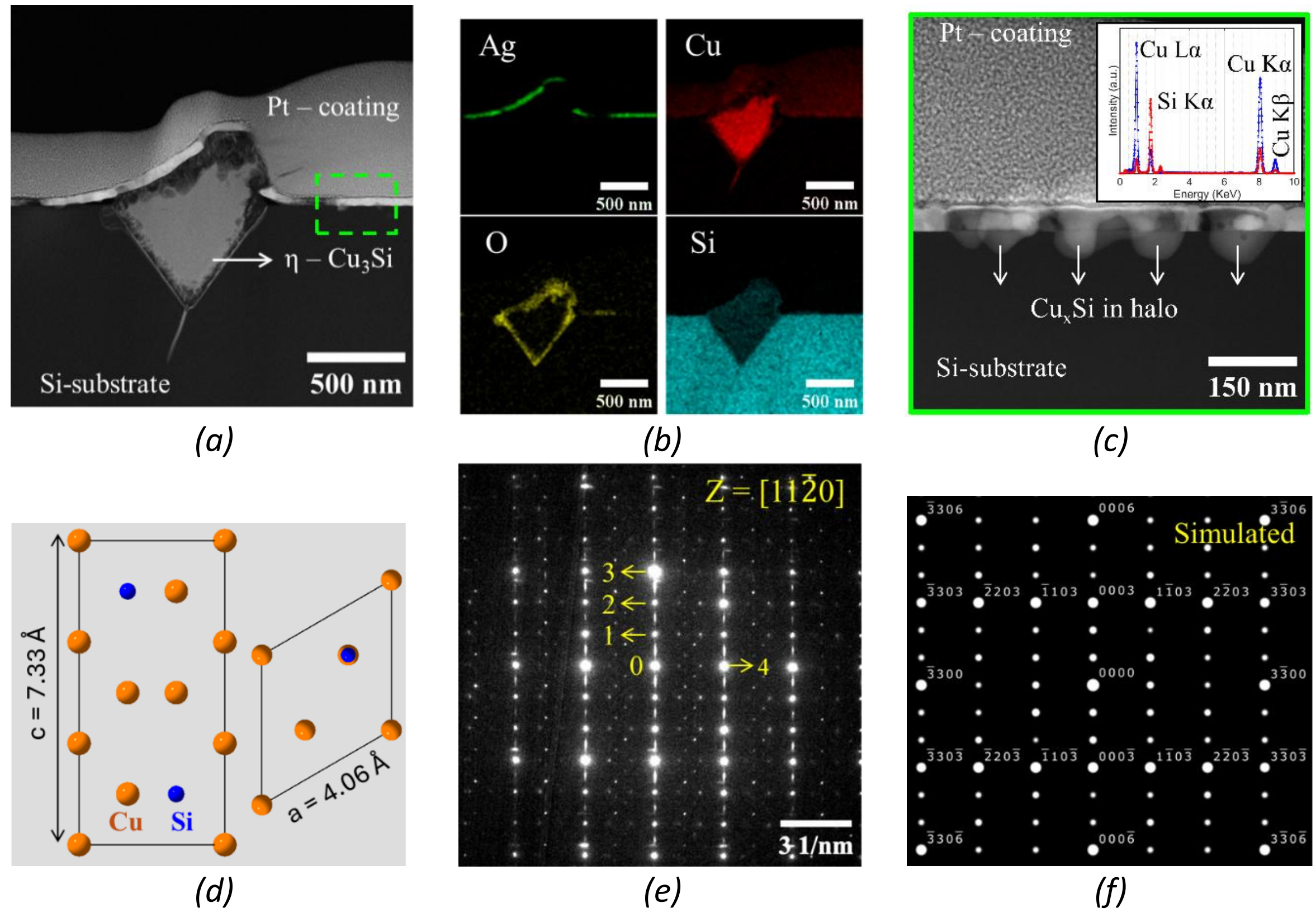

***Figure 3:*** *(a) HAADF-STEM image showing a cross-sectional view of the bright particle formed by the film-substrate reaction. (b) STEM-EDS maps of Cu, Ag, Si and O. (c) HAADF-STEM image of tiny particles in the halo adjacent to the central copper silicide. Inset: STEM-EDS spectra of both types of silicide particles comparing their Cu:Si ratios; blue peaks are from the large silicide with a stoichiometry $Cu_{79}Si_{21}$, whereas the red peaks are from the tiny particles in (c) which have ~70-75 at% Si. (d) Unit cell of η-$Cu_3Si$ viewed along (left)* ***a*** *and (right)* ***c*** *axes. Lattice parameters, Wyckoff positions and crystal symmetry are taken from Wen and Spaepen* [70]. *(e) ED pattern from the central silicide (η-$Cu_3Si$) particle in Figure 3(a) along the* $[11\bar{2}0]$ *zone. Reflections corresponding to labels 1 to 4 correspond to (0001), (0002), (0003) and* $(1\bar{1}00)$ *planes of the trigonal η-$Cu_3Si$ structure. (f) Simulated EDP corresponding to the unit cell shown in (a).*

As revealed in the cross-sectional image presented in Figure 3(a), the central silicide particle grows both in the lateral direction (parallel to the plane of the film) as well as in directions normal to it. It penetrates the silicon surface and grows inward to a depth of ~550 nm while maintaining a characteristic V-type shape and also toward and beyond the top surface of the film. This phase starts to form along the film-substrate interface in the aperture, and given enough time at an elevated temperature, fills the aperture completely by spreading laterally. Its depth-wise growth leads to the characteristic morphology. The angle between the inclined interface between $Cu_3Si$ and the top surface of the (100) Si substrate, as measured from Figure 2(a), is about 54°. This suggests that the growth indeed proceeds along {111} planes of the Si substrate. The geometry of this specific mode of growth constrains the width to depth ratio of the particle to be 1.42; the observed value of about 1.4 therefore provides support for this mode of growth. Viale *et al.* [79] showed that the etching of $Cu_3Si$ particles grown on Si wafers left behind pits with an inverted pyramid shape, indicting that growth takes place along certain crystallographic planes of silicon. Li *et al.* [80] reported that the growth of $Cu_3Si$ takes place preferentially along Si-$\{111\}$ planes giving rise to characteristic V-type morphology. Similar observations have also been reported by other studies [81,82].

### *3.3 Isothermal growth kinetics*

During post-deposition annealing, Cu in the metastable film reacts with the exposed Si substrate in the aperture to form $Cu_3Si$. Its growth leads to the depletion of Cu from the adjacent film, eventually forming the Ag-rich halo surrounding the silicide product phase. As observed in Figures 2(b) and 3(c), finer copper silicides (features with a darker contrast in Figure 2(b)) also form inside the halo, albeit to a lesser extent. Based on our experiments, the onset temperature for the silicide formation reaction is around 180 °C when a discontinuous halo appears to form after 4 h of annealing (Figure S6(a)). Silicide formation, accompanied by a narrow and continuous halo, is more clearly observed after 0.5 h at 200 °C (Figure S6(b)). This agrees well with studies that report [66,67] that $\sim$200 °C is the critical temperature above which $Cu_3Si$ forms during annealing of elemental Cu films grown on Si-wafers. Since the reaction becomes reasonably faster and development of the halo is more pronounced only beyond 300 °C, their growth kinetics is investigated by carrying out annealing experiments for different times at 300 °C and 350 °C.

Figure 4 shows the sequence of microstructural evolution at 300 °C and 350 °C for increasing times from 0.5 h to 3 hr. In these images, we can observe a bright, blocky $Cu_3Si$ particle in the center which is surrounded by finer copper silicides (grey contrast) with an irregular morphology. The sizes of both the $Cu_3Si$ particle and the halo increase with annealing time and temperature, indicating their correlation. The halo growth is a consequence of interdiffusion of Cu and Ag atoms triggered by formation of $Cu_3Si$, with Cu being towards the aperture and Ag away from it. Thus, these images establish the thermally activated nature of the process that depends on species transport through the film near the aperture.

The variation in the width of the $\eta$-$Cu_3Si$ particle and that of the halo (defined as the distance from the edge of the silicide to the halo/bulk film boundary indicated by the dashed circles in Figure 4) as a function of annealing time at the two different temperatures are plotted in Figures 5(a) and (b). For statistically reliable estimates, mean widths are obtained by averaging results from multiple apertures for each annealing condition. When the experimental data is fit to a simple two-parameter power law relation of the type $\lambda = kt^n$ ($\lambda$ is the mean length scale, $t$ is the annealing time, $n$ is the growth exponent and $k$ is a pre-factor which depends on temperature and material properties), the growth exponents $n^P$ and $n^H$ for the product silicide and the halo for both temperatures are found to be 0.3 and 0.46, respectively. For phase formation by solid-state reaction between metallic thin films and Si, the exponent is reported to be 0.5 and 1.0 for diffusion- and reaction-controlled kinetics, respectively [50]. The deviation of the growth exponents from these expected values will be explored further in **Section 3.5** where we shall present a semi-analytical model for the growth kinetics based on species and flux balance.

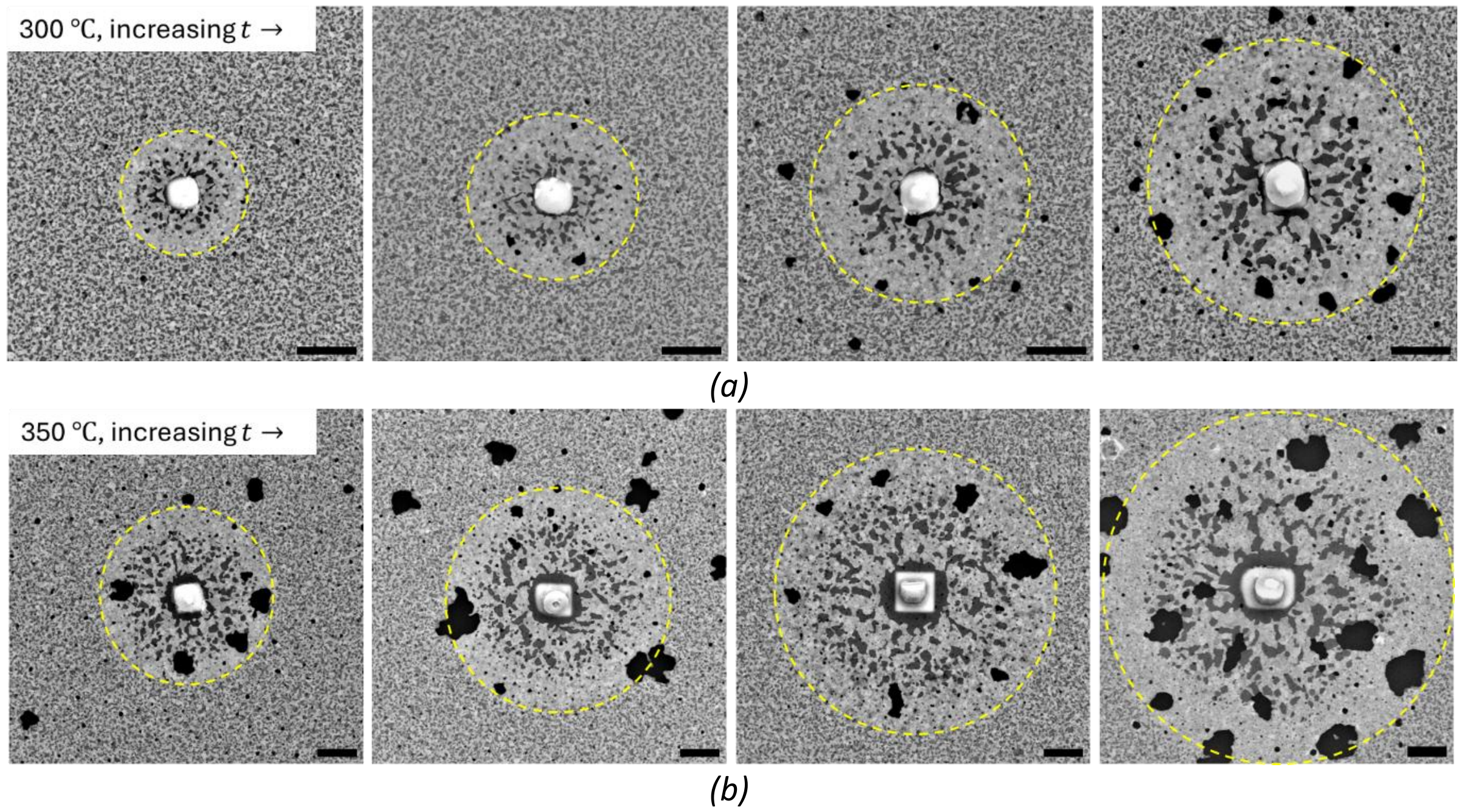

***Figure 4:** Evolution of $Cu_3Si$ and halo structure at different annealing conditions for annealing temperatures of (a) 300 ℃ and (b) 350 ℃, respectively. Increasing times from left to right correspond to annealing times of 0.5, 1, 2 and 3 h.* The scale bars represent 2 μm.

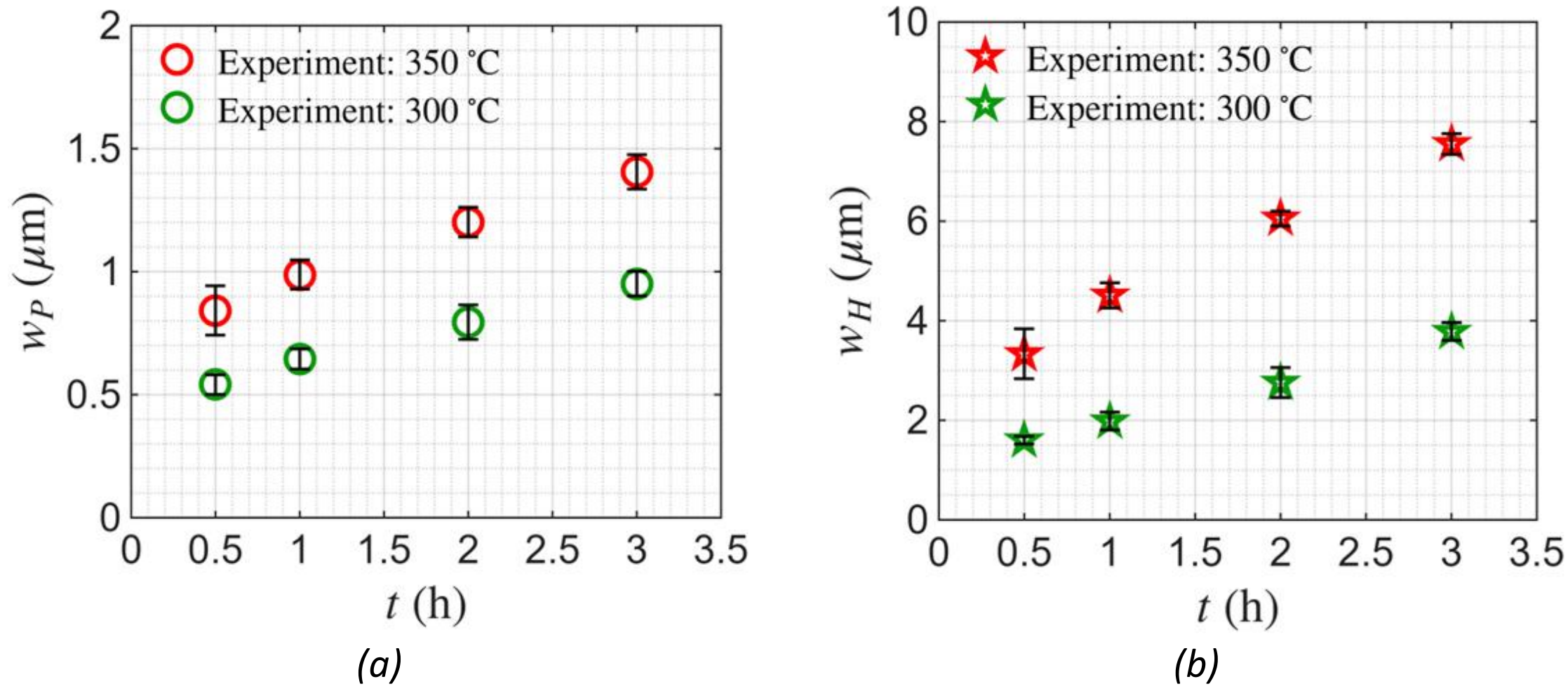

***Figure 5:** The widths of the (a) product $Cu_3Si$ and (b) the halo measured from SEM-BSE micrographs of the films annealed at different conditions*.

### *3.4 Microstructural changes in the bulk film*

Metastable Ag-Cu films progressively decompose into Ag-rich and Cu-rich domains via spinodal decomposition during post-deposition annealing which subsequently coarsen with time [83–85]. The extent or progress of the decomposition process is determined by tracking the change in the lattice parameters of the Ag- and Cu-rich phases as a function of the annealing time. Figure 6 shows the GIXRD profiles obtained from as-deposited and annealed films at selected annealing conditions. In the as-deposited state, there is a broad and intense peak at 39.6° originating from {111} planes of the undecomposed FCC solid solution with a lattice parameter of 3.94 Å (those of pure Ag and Cu are 4.07 and 3.59 Å, respectively). At increased annealing temperature and/or time, multiple and sharper peaks corresponding to Ag- and Cu-rich phases appear, indicating the onset and progress of phase separation. Further, phase separation occurs at temperatures as low as 150 °C, which is ~50 °C lower than the critical temperature for the onset of silicide reaction. This clearly establishes that phase separation happens prior to the formation $Cu_3Si$ and halo structure.

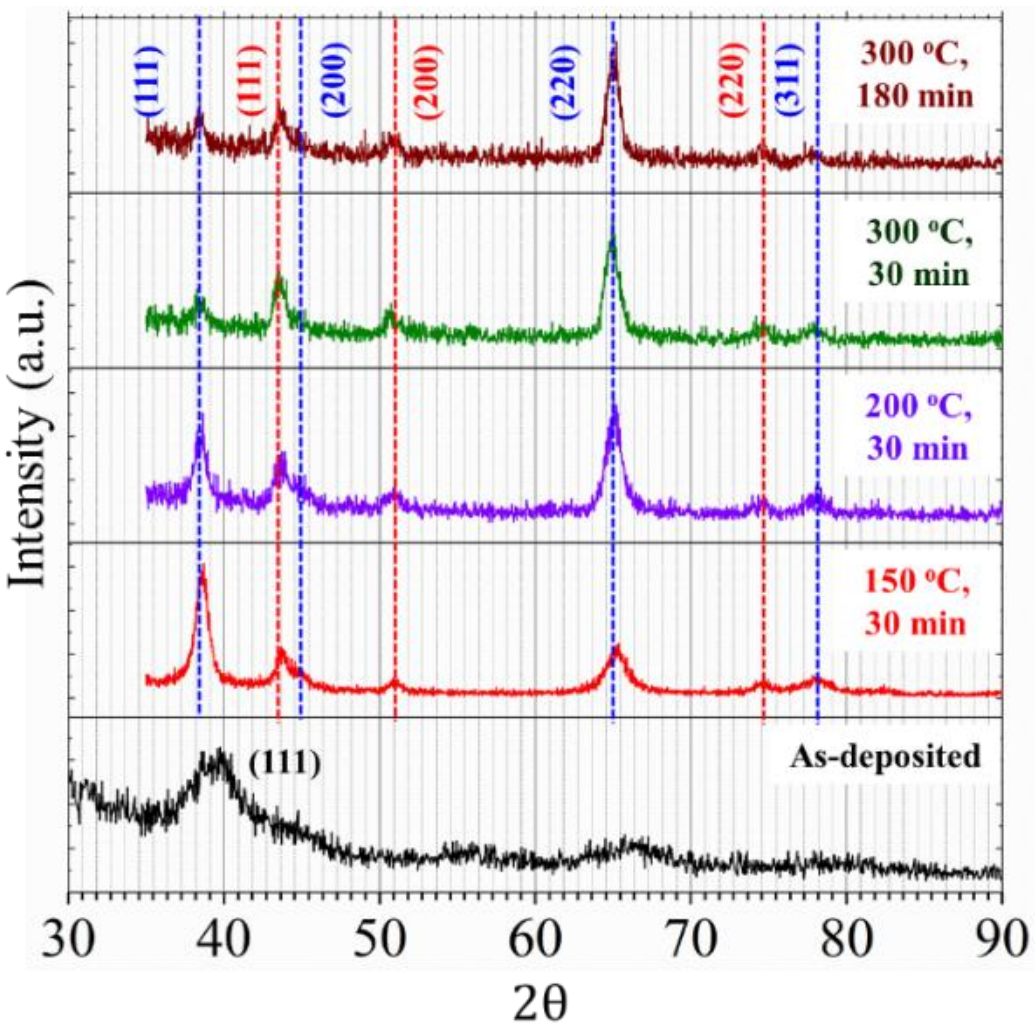

***Figure 6:*** *GIXRD patterns for films annealed at different temperatures and times. Vertical blue and red dashed lines represent Ag-rich and Cu-rich phases, respectively.*

The effect of temperature and time of annealing on the phase separation in the bulk film at 200 °C and 350 °C is illustrated in Figure 7 where the bright and dark regions correspond to Ag- and Cu-rich domains, respectively; micrographs for intermediate temperatures are provided in Figure S7. In the bulk film, the Cu-rich phase appears as isolated domains embedded in an Ag-rich matrix, and they become larger at higher annealing temperatures. The amount or fraction of Cu-rich domains and their characteristic length scale are determined independently using the two different methods (*viz.*, ACF and CLD) discussed in **Section 2.3**. Figure 8(a) presents radial ACF line profiles obtained by circular averaging of the 2d ACF map for different annealing conditions. The first zero of the radial profile is taken as the characteristic microstructural length scale [86], and its variation with time and temperature is shown in Figure 8(b). Alternatively, we also estimated the length scale from CLDs. Since CLDs can be obtained for different directions in the image, they provide additional information regarding preferential alignment of domains; Figure 8(c) shows an absence of any such alignment. The time-temperature variation of the mean length obtained from CLD analysis is presented in Figure 8(d), which shows a fair agreement with the same determined by the ACF method.

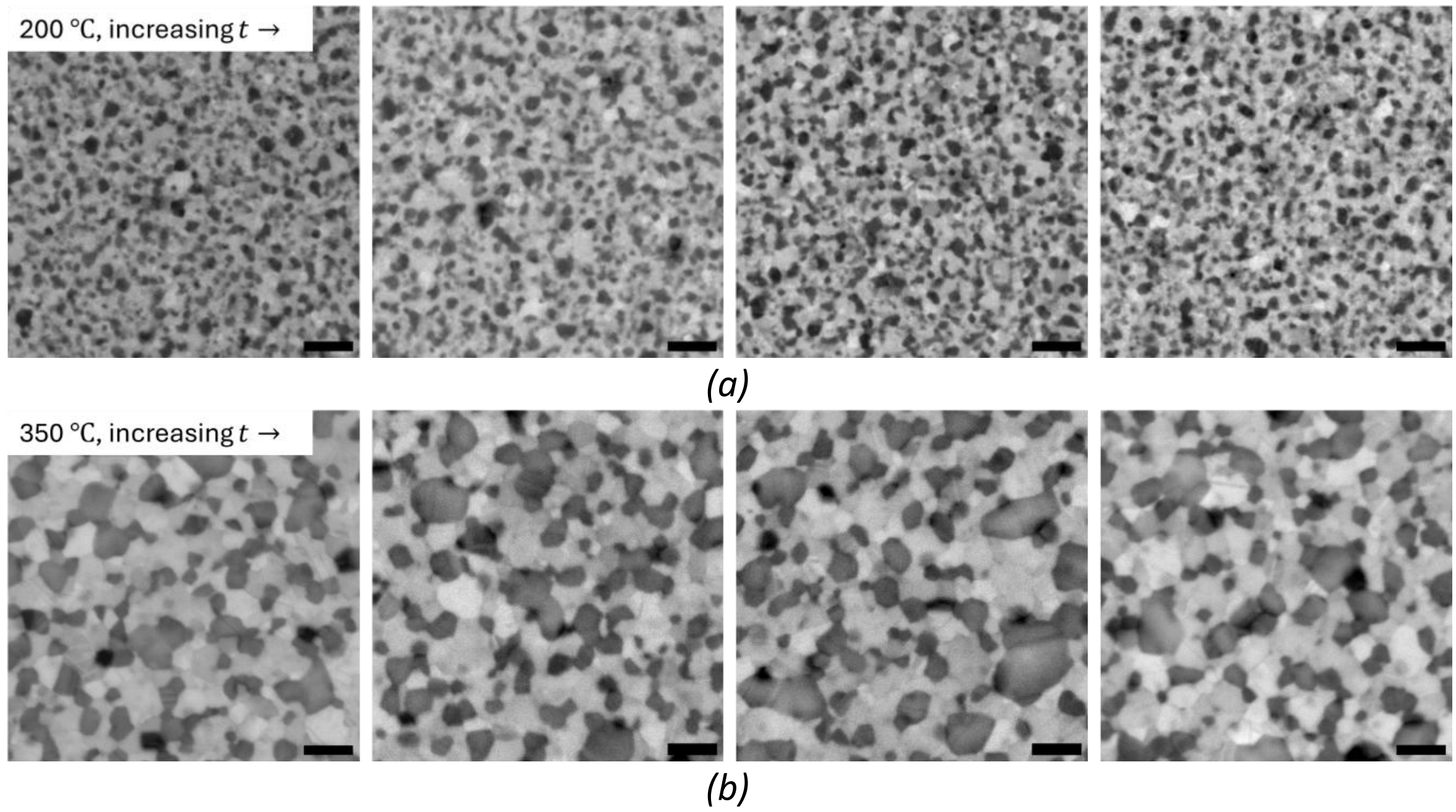

***Figure 7:*** *Representative bulk film microstructures showing Ag-rich (bight) and Cu-rich (dark) domains formed in films when annealed at 200 °C (a – d), and 350 °C (e – h) for different times. Increasing times from left to right correspond to annealing times of 0.5 h, 1 h, 2h and 3h. The scale bar corresponds to 250 nm.*

During annealing at a given temperature, Cu-rich domains develop to a stable size soon after the initial separation and they resist further coarsening with increasing time. This limiting size of the domains increases from ∼50 nm to ∼110 nm as the annealing temperature is increased from 200 °C to 350 °C. A similar phenomenon was also observed in a previous study [83] on spinodal decomposition of metastable Ag-Cu thin films using *in situ* electron microscopy, but quantitative details or explanatory mechanisms were not presented. The current situation differs from coarsening of domains produced by spinodal decomposition or precipitate coarsening in one crucial aspect: in these latter cases, the phases undergoing coarsening are inside a single grain, whereas here, each domain may span across several grains. Since coarsening depends on diffusion flux created by the Gibbs-Thomson effect, diffusion across and along grain boundaries (GBs) may influence its rate. In another regard, since domain coarsening is a capillarity driven phenomenon, it resembles grain growth, with domain boundary energy playing an analogous role as the grain boundary (GB) energy. Similar to our observations, studies on grain growth kinetics in thin films too show that grain size increases more significantly with annealing temperature than with annealing time [87–89]. Migrating GBs and domain boundaries may interact, both through their energies and mobilities, and the two processes may exhibit correlative effects. How this affects the microstructural evolution in thin films needs to be investigated further.

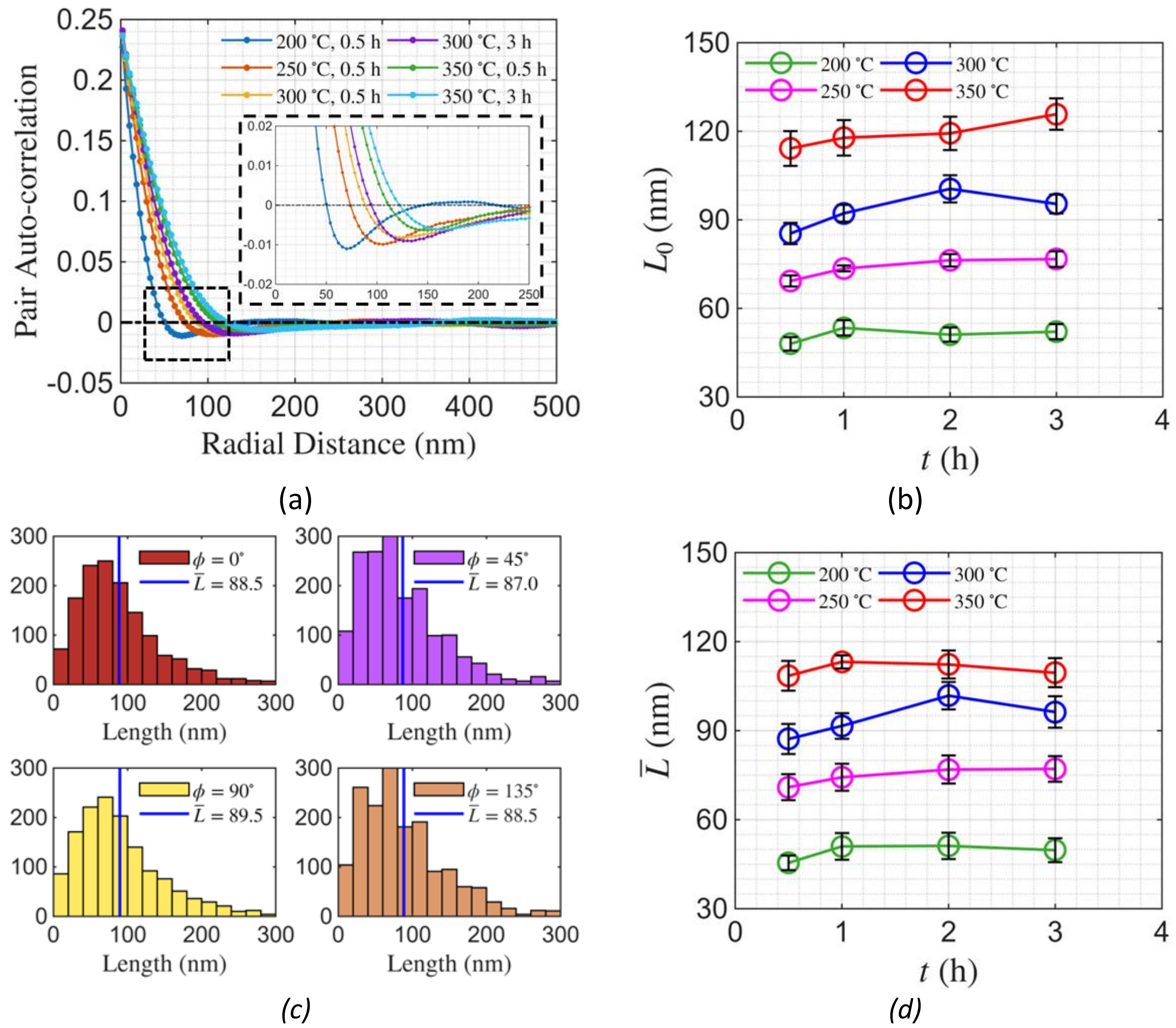

***Figure 8:*** *(a) Radial ACF profiles (b) time-temperature dependence of the length scale (the first zero or radial ACF). (c) Frequency distribution of chord lengths along 0°, 45°, 90° and 135° to the horizontal. (d) The mean chord length of the CLD as a function of annealing time at different temperatures. Both metrics used to quantify the size of Cu-rich domains in phase separated film are equivalent.*

### *3.5 Reaction-diffusion model of silicide growth and evolution of the halo*

Our experimental results demonstrate silicide formation by film-substrate reaction in and around milled apertures. The central silicide particle is always accompanied by a surrounding Ag-rich halo, which forms as a direct consequence of the depletion of Cu from the adjacent film as copper is transported to the reaction front by diffusion. To determine the time-dependence of the two relevant length scales (*viz.*, widths of the $Cu_3Si$ particle and the halo), we now present a kinetic model of this reaction-diffusion process. We develop the model by considering diffusion of Cu through the adjacent film to be the rate controlling step for growth.

***The growth model****:* Let us consider a single central silicide particle growing both laterally into the film and depth-wise into the substrate; the formation of other silicide particles are ignored. As shown schematically in Figure 9(a), we consider a Cu-depleted zone in the immediate vicinity of the reaction product, and following that, the film consists of Cu-rich domains embedded in an Ag-rich matrix due to partial or complete phase separation. Since the Ag-rich phase has very limited solubility of Cu, the supply of Cu to the reaction front takes place primarily through the GBs of Ag, both because they can accommodate higher Cu-concentrations ($C_{\mathrm{Cu}}^{GB} \gg C_{\mathrm{Cu}}^{(Ag)}$) due to segregation [90,91] and also because they offer a high-diffusivity path compared to the lattice ($D_{\mathrm{Cu}}^{GB} \gg D_{\mathrm{Cu}}^{L}$). Additionally, we simplify the problem by coarse-graining these microscopic details and assuming the diffusive transport to be driven by an effective or homogenized Cu concentration gradient through a medium. This is shown in Figure 9(b) where, similar to the Zener profile for precipitate growth, a linear gradient has been assumed between the far-field field with an average Cu composition and the reaction front with a fixed Cu concentration.

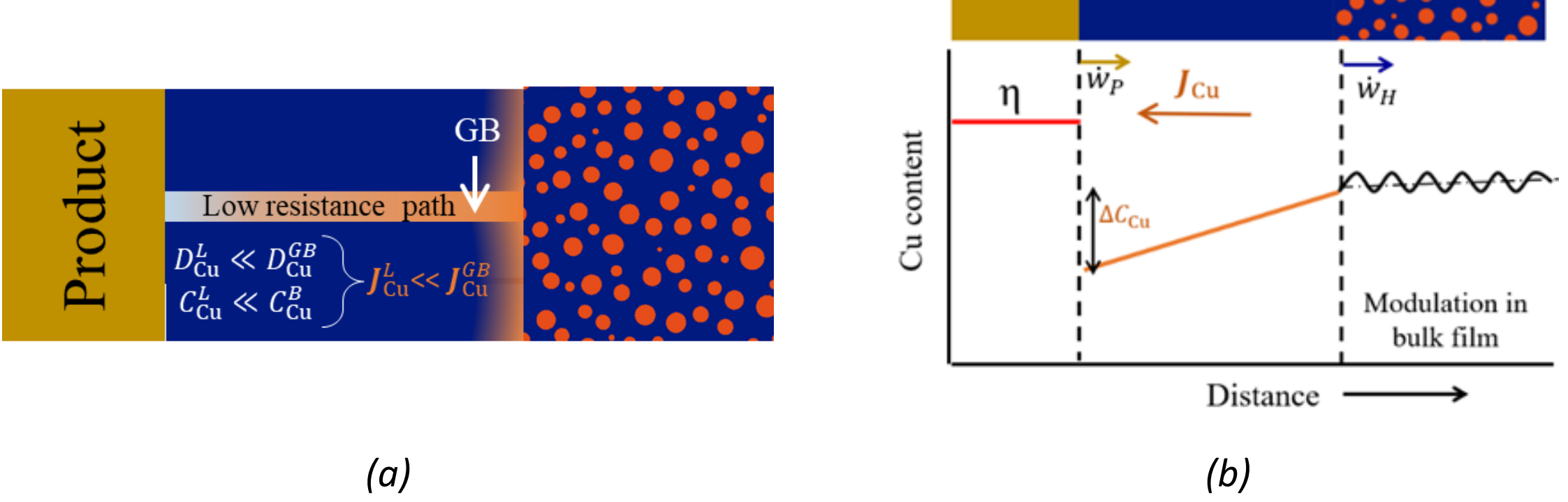

*(a)* *(b)*

***Figure 9:*** *(a) Schematic top view with microscopic details of the halo (largely made up of Ag-rich grains and grain boundaries) and the bulk film (Cu-rich domains embedded in an Ag-rich matrix. (b) Variation of the Cu-content obtained by smearing out the microscopic details. A constant composition in the product, a linear profile in the halo and small deviations from the mean composition in the far field are assumed.*

Figures 10(a) and 10(b) present the top and cross-sectional views, respectively, of this simplified model. The symbols $w_P$ and $w_H$ represent half-widths of the reaction product ($P$) and halo ($H$), respectively, at an instant of time $t$, $h_F$ is the film thickness, $\theta$ is the observed characteristic angle of growth (54.7°) and $\boldsymbol{J}_{\mathrm{Cu}}$ represents the flux of Cu through the halo; incremental increases in $w_P$ and $w_H$ in a time increment $dt$ are also indicted.

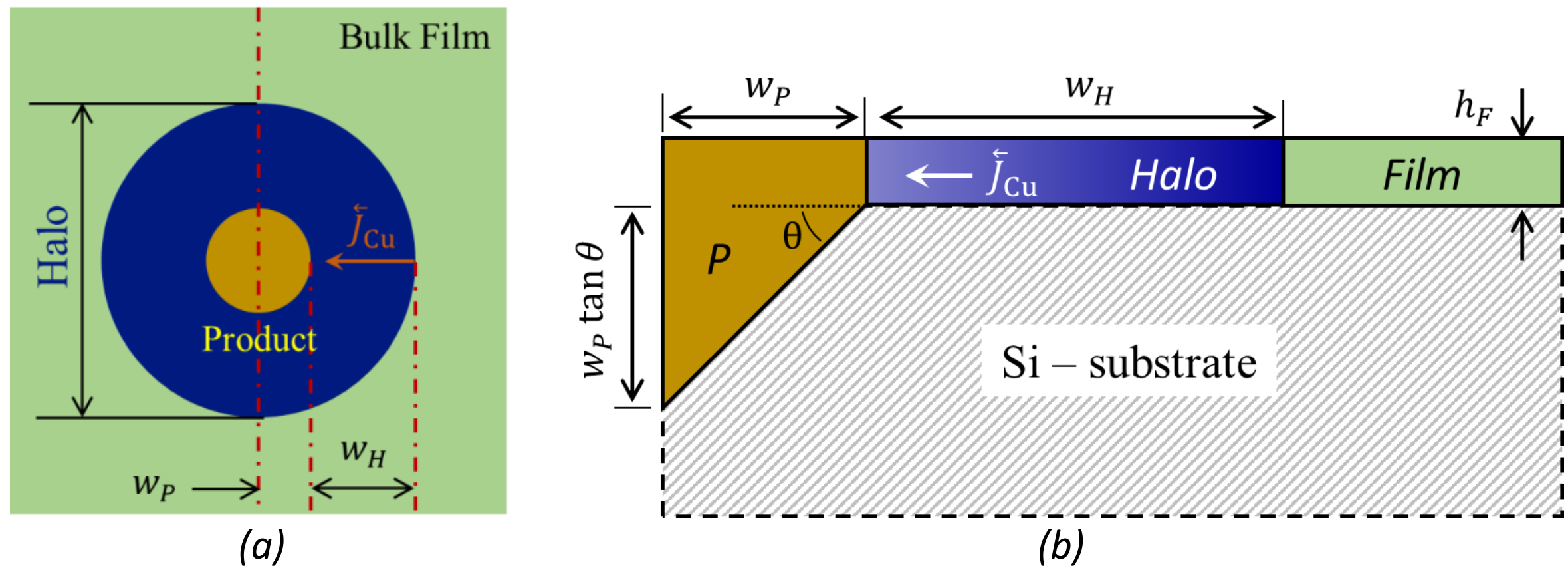

*(a)* *(b)*

***Figure 10:*** *Schematic representation of the geometry, processes and variables involved in the formation of the reaction product and the halo. (a) Top view and (b) cross-sectional view.*

***Species balance***: Since the growth of $Cu_3Si$ and its associated halo are interrelated processes contingent upon the supply of Cu-atoms, we can arrive at a correlation between $w_P$ and $w_H$ from the conservation of Cu. Let us consider a general case where a film of volume $V^F$ gets replaced by a product phase of volume $V^P$ and a halo of volume $V^H$; note that all these volumes increase with increasing time $t$. We express the species balance of Cu atoms valid for all times as follows:

$$n_{\mathrm{Cu}}^F = n_{\mathrm{Cu}}^P + n_{\mathrm{Cu}}^H \tag{1}$$

where $n_{\mathrm{Cu}}^i$ ($i = F$, $P$ or $H$) are the number of Cu atoms in the film, the product and the halo, respectively. Note that $n_{\mathrm{Cu}}^i = V^i \rho_{\mathrm{Cu}}^i X_{\mathrm{Cu}}^i$ where $\rho_{\mathrm{Cu}}^i = N_A/\Omega^i$ is the number of Cu atoms per unit volume (the number density) with $\Omega^i$ being the molar volume of $i$ and $N_A$ the Avogadro's number. The mole fraction of copper $X_{\mathrm{Cu}}^i$ in the initial film and the resulting halo in the phase-separated film can be obtained by averaging over the Cu-rich $\alpha$ and Ag-rich $\beta$ solid solution phases (which, for an undecomposed or single-phase material would correspond to its value in that phase). We introduce factors $s^i$ to express the $X_{\mathrm{Cu}}^i$ in terms of the equilibrium Cu mole fraction in the Cu-rich $\alpha$ phase:

$$X_{\mathrm{Cu}}^i = s^i X_{\mathrm{Cu}}^\alpha, i = F, P, \beta, H. \tag{2}$$

Since the volume of the film that gets replaced by the reaction product and the halo can be assumed to be a circular cylinder of radius $(w_P + w_H)$ and height $h_F$; thus, we have $V^F = \pi(w_P + w_H)^2 h_F$, and the total number of Cu atoms in the initial film is given by:

$$n_{\mathrm{Cu}}^F = \pi(w_P + w_H)^2 h_F \rho_a^F s^F X_{\mathrm{Cu}}^\alpha \tag{3}$$

As evident from Figure 10(b), the volume of the reaction product can be split into two parts: (1) a top cylindrical part of radius $w_P$ and height $h_F$ has the volume $V_1 = \pi w_P^2 h_F$ and (2) a bottom conical part with a base radius of $w_P$ and height $w_P \tan\theta$, so $V_2 = \frac{1}{3}\pi w_P^2 (w_P \tan\theta)$. The angle $\theta$ made by the product with the top surface of the substrate originates from the observation that the product grows along the {111} planes of Si-(100) substrate, thus giving rise to the characteristic V-type morphology with $\theta = 54.7°$.

Therefore, $V^P = V_1 + V_2 = \pi w_P^2 h_F + 1/3\, \pi w_P^2 (w_P \tan\theta)$, and number of copper atoms in the product is obtained as

$$n_{\mathrm{Cu}}^P = V^P \rho_a^P X_{\mathrm{Cu}}^P = \pi \left(\frac{h_F}{w_P} + \frac{1}{3}\tan\theta\right) w_P^3 \rho_a^P s^P X_{\mathrm{Cu}}^\alpha. \tag{4}$$

The halo constitutes of an annular region between the product and the far-field film, and its volume is given by $V^H = \pi[(w_P + w_H)^2 - w_P^2] h_F$, thereby yielding the total number of Cu atoms in the halo as:

$$n_{\mathrm{Cu}}^H = \pi(w_H^2 + 2 w_H w_P) h_F \rho_a^H s^H X_{\mathrm{Cu}}^\alpha \tag{5}$$

Inserting Eqs. 3-5 into Eq. 2 and rearranging terms, we obtain the following quadratic equation for $w_H$ in terms of $w_P$:

$$w_H^2 + 2 w_P w_H - \left[\left(\frac{\tan\theta}{3h_F}\,\frac{\rho_a^P s^P}{\rho_a^F s^F - \rho_a^H s^H}\right) w_P^3 - \left(\frac{\rho_a^F - \rho_a^P s^P}{\rho_a^F s^F - \rho_a^H s^H}\right) w_P^2\right] = 0 \tag{6}$$

Ignoring the negative root of this equation, the physically meaningful solution for $w_H$ is:

$$w_H = -w_P + w_P \left[ \left( \frac{\tan\theta}{3h_F} \frac{\rho_a^P s^P}{\rho_a^F s^F - \rho_a^H s^H} \right) w_P + \left( \frac{\rho_a^P s^P - \rho_a^H s^H}{\rho_a^F s^F - \rho_a^H s^H} \right) \right]^{\frac{1}{2}} \tag{7}$$

Before proceeding further, it is worth discussing the limiting values for $\rho_a^i$, $s^i$ and $X_{\mathrm{Cu}}^i$. For the product, $\rho_a^P$'s can be obtained from its molar volume, and for the film and the halo, $\rho_a^i$ can be computed from the molar volumes of $\alpha$ and $\beta$ in a phase-averaged manner; all these are of similar orders. For the present case, we also have $X_{\mathrm{Cu}}^\alpha > X_{\mathrm{Cu}}^P > X_{\mathrm{Cu}}^F > X_{\mathrm{Cu}}^H > X_{\mathrm{Cu}}^\beta \Longrightarrow 1 > s^P > s^F > s^H > s^\beta$ (again, phase averaged values used for the film and the halo). For different initial film composition and product stoichiometry, corresponding values can be used directly. For the halo, the average $X_{\mathrm{Cu}}^H$ can be measured experimentally, or, for the linear composition gradient shown in Figure 9(b), it is the average of the far-field film composition ($X_{\mathrm{Cu}}^F$) and that at the reaction front ($X_{\mathrm{Cu}}^{RF}$); the latter can be taken as close to zero for the Cu-$Cu_3Si$ equilibrium. We use the following values for relevant parameters in our calculations: $X_{\mathrm{Cu}}^\alpha$= 0.95, $X_{\mathrm{Cu}}^P$= 0.75, $X_{\mathrm{Cu}}^F$= 0.5, $X_{\mathrm{Cu}}^H$= 0.25, $X_{\mathrm{Cu}}^\beta$= 0.05, $\rho_a^P$= 8.6 × $10^{10}$, $\rho_a^F$ = 6.72 × $10^{10}$, $\rho_a^H$= 1.84 × $10^{10}$ (at/μm$^3$).

We note that in the model development, we neglected the presence of pores and smaller secondary silicide particles in the halo region. To a first approximation, the neglect of the former (*i.e.*, the pore volume) leads to an overestimate of $w_H$ by a small factor because pores can be thought of as a zero-Cu phase, making more Cu available for the growth of the reaction product (that is, they act as a fictitious Cu-source). On the other hand, secondary silicides trap Cu and reduce the amount of Cu available for the product. This leads to an underestimate of $w_H$ by the model. Consideration of these two aspects will introduce two correction pre-factors (one greater than and the other less than unity, respectively) in the model. Importantly, however, these do not affect the power law exponent in Eq. 7 and the subsequent analysis.

***Predictions from species balance*:** $w_H$ calculated from Eq. (7) is plotted in Figure 11(a) as a function of $w_P$. On this plot, we also superimpose the experimentally measured ($w_P$, $w_H$) pairs at two different temperatures. Despite the simplifying assumptions of the model, the observed data appear to agree very well with the prediction without requiring any additional fitting factor.

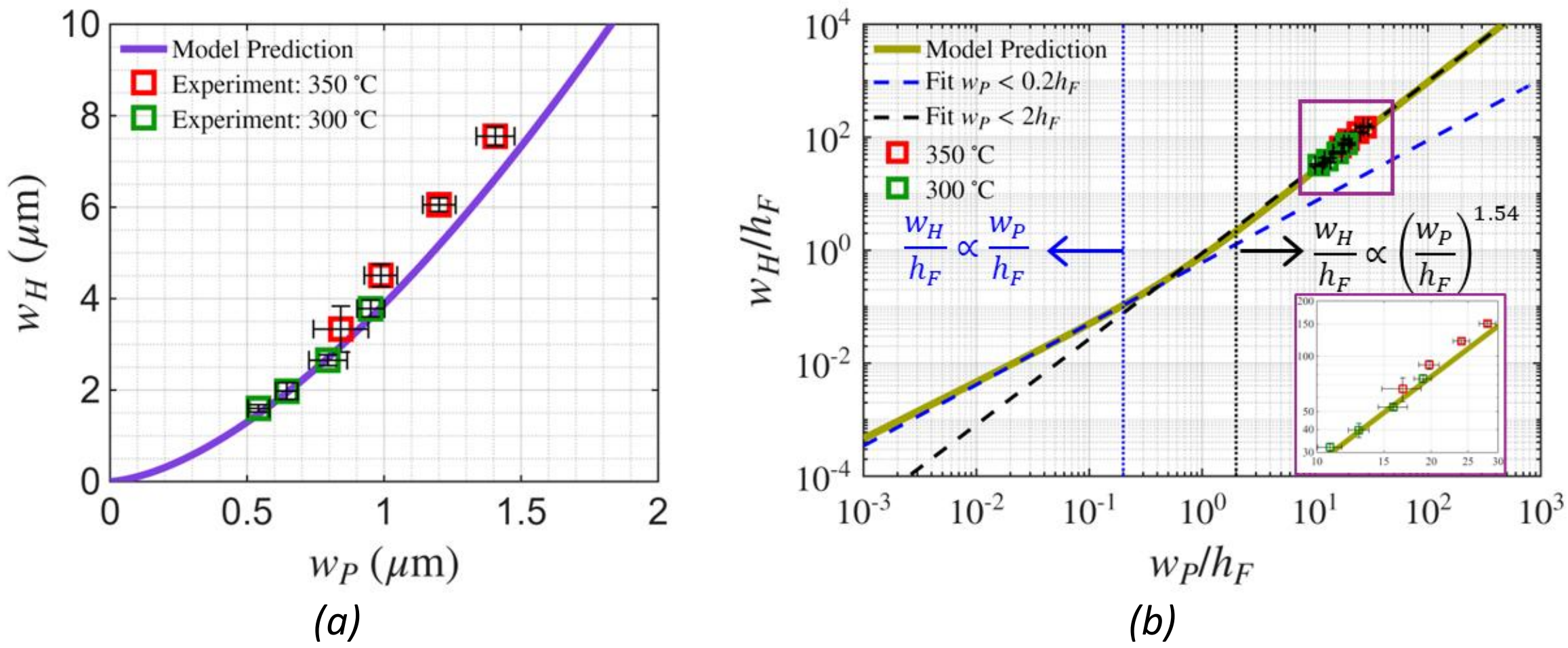

***Figure 11:*** *(a) Variation in the width of halo ($w_H$) as a function of the width of the product ($w_P$). The curves are model predictions (Eq. (7)), and green and red markers are the experimental data obtained at 300 °C and 350 °C, respectively. (b) The normalized $w_H$-$w_P$ relation obtained by scaling the quantities by the film thickness $h_F$. The log-log plot of reveals two limiting regimes.*

To extend the applicability of the model predictions further, we normalize the lengths $w_P$ and $w_H$ by the film thickness $h_F$, and plot it in Figure 11(b). To explore the existence of a power-law relationship of the type $w_H \propto w_P^m$, the prediction is plotted on a log-log scale in Figure 11(b). Again, the experimental data fit the scaled variables very well in these plots; see inset in Figure 11(b). Additionally, Figure 11(b) reveals that although no universal power-law relation is applicable over the entire range of values, there exist two distinct power-law regimes for $w_P/h_F \leq 0.2$ and $w_P/h_F \geq 2$. The current experiments fall into the latter case with an index of 1.54 (the black dashed line), while the former exhibits an almost-linear correlation with an index of $\sim$1 (the blue dashed line). Note that the former is very close to the value of 3/2 which is the power of the leading order term ($w_P \times w_P^{1/2}$) in Eq. 7. From the definitions of the two volumes $V_1$ and $V_2$ (introduced before Eq. 4) that make up $V^P$, we get $w_P/h_F = 2V_2/V_1$; thus, a value of 2 for $w_P/h_F$ corresponds to the case when these two volumes are equal. Variation of the relative dimensions of the product (in-film and in-substrate) with film thickness is shown schematically in Figure S8. For relatively thin films ($w_P/h_F \gg 2$), most of the product grows into the substrate, which requires a larger halo region from which solute is to be supplied, thereby giving rise to a higher index. On the other hand, most part of the product in thicker films ($w_P/h_F \ll 0.2$) remains within the film itself, and the halo width exhibits a linear power-law relationship with the product width.

***Flux balance****:* For the growth of the product, Cu atoms must be supplied at the growing reaction front ($RF$) that is made up of two different interfaces: product-film ($PF$) and product-substrate ($PS$). Drawing out Cu atoms from the Ag-Cu film away from the interface eventually would lead to the formation of the Cu-depleted (*i.e.*, Ag-rich) halo region. Note that in general, dimensionality of solute transport is 2d while that of the growth can be either 2d (thick film limit) or 3d (thin film limit). Since the growth of the product takes place with two different jump conditions for the Cu concentration at $PF$ and $PS$ interfaces that constitute the $RF$, to obtain the growth velocity, we need to modify the classical Stefan moving boundary condition. This is schematically represented in Figure 12.

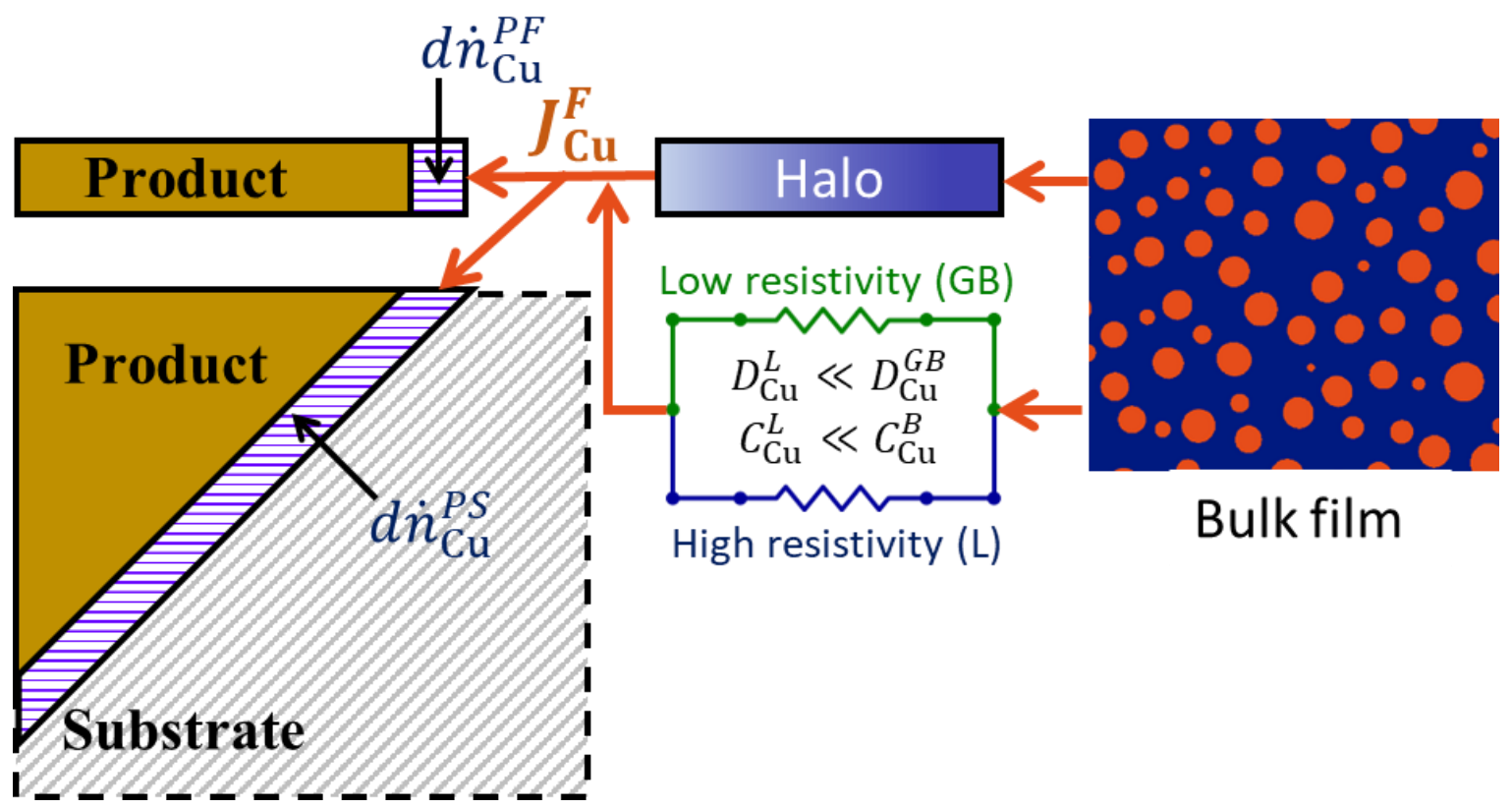

***Figure 12:*** *Sketch showing the modified Stefan balance condition. The total amount of solute supplied from the bulk of the film is redistributed to (i) flat cylindrical part at top and (ii) conical part at bottom of the product phase with two different interfacial jump conditions.*

The interface species balance equation is expressed by equating (1) the net *rate of transport* of Cu atoms through the film by diffusion with the (2) the *incremental rate of accumulation* of Cu atoms in the product. Since there is only one source of Cu atoms, namely, the film surrounding the product, the first term is simply the magnitude of the diffusion flux of Cu atoms in the film, $J^F_{\mathrm{Cu}}$, times the cross-sectional area of the $PF$ interface, $A_{PF} = 2\pi w_P h_F$. Since the incoming flux gets split into two parts (see Figure 11(b) for a schematic illustration), the second term is the sum of two terms representing the rates of solute accumulation in two parts of the product volume (composition jump $\Delta C^i_{\mathrm{Cu}}$ at each of the two interfaces times the corresponding volume $V_i$). Therefore, we can thus write:

$$\frac{d}{dt}\left(\Delta C^{PF}_{\mathrm{Cu}} V_1 + \Delta C^{PS}_{\mathrm{Cu}} V_2\right) = J^F_{\mathrm{Cu}} \times A_{PF} \tag{8}$$

Note that $\Delta C^{PF}_{\mathrm{Cu}} = C^P_{\mathrm{Cu}} - C^{PF}_{\mathrm{Cu}}$, $\Delta C^{PS}_{\mathrm{Cu}} = C^P_{\mathrm{Cu}} - C^{PS}_{\mathrm{Cu}} \approx C^P_{\mathrm{Cu}}$ (since the solubility of Cu in Si is very low), and the volumes $V_1$ and $V_2$ are defined as before: $V_1 = \pi w_P^2 h_F$, $V_2 = (\pi/3)\tan\theta\, w_P^3$. To estimate the diffusion flux magnitude $J^F_{\mathrm{Cu}}$, we note that as shown in Figure 9(b), the Cu concentration profile is assumed to be linear between the $PF$ interface with Cu concentration $C^{PF}_{\mathrm{Cu}}$ and the far-field film with a Cu concentration of $C^F_{\mathrm{Cu}}$. Since this change in concentration takes place over a distance $w_H$, the diffusion flux magnitude is given as:

$$J^F_{\mathrm{Cu}} \cong D_{\mathrm{Cu}} \frac{C^F_{\mathrm{Cu}} - C^{PF}_{\mathrm{Cu}}}{w_H} \tag{9}$$

where $D_{\mathrm{Cu}}$ is the effective or homogenized diffusivity of Cu in the film (which we approximate to be composition-independent).

We insert Eq. (9) into Eq. (8) and use the definitions of $V_i$, $A_i$ and $\Delta C^i_{\mathrm{Cu}}$, to obtain the following governing equation for the diffusion-controlled growth rate of $Cu_3Si$:

$$\frac{dw_P}{dt} = \frac{D_{\mathrm{Cu}}\left(C^F_{\mathrm{Cu}} - C^{PF}_{\mathrm{Cu}}\right)}{w_H\left[\left(C^P_{\mathrm{Cu}} - C^{PF}_{\mathrm{Cu}}\right) + \tan\theta\, C^P_{\mathrm{Cu}}\left(\frac{w_P}{2h_F}\right)\right]} \tag{10}$$

The expression for $w_H$ in terms of $w_P$ from Eq. (7) can be inserted in Eq. (10), leading to a first order ordinary differential equation (ODE) for $w_P$. Note that the presence of the $w_H w_P$ term in the denominator of the RHS, in which $w_H$ itself has a functional dependence on $w_P$ (Eq. 7), makes it challenging to obtain an analytical solution without further simplifications of the terms. Therefore, we proceed to solve Eq. 10 numerically using the NDSOLVE routine in Mathematica. At the *PF* interface we expect to have almost pure Ag, so we assumed $X_{\mathrm{Cu}}^{PF} = 0.05$ for the calculations.

***Predictions from the growth model***: Except diffusivity $D_{\mathrm{Cu}}$, values of all parameters in Eq. (10) can be estimated either from our experimental observations or from system thermodynamics. We therefore used $D_{\mathrm{Cu}}$ as a parameter while solving for $w_P(t)$. Note that a zero-width initial condition (IC) cannot be used for solving the ODE as the RHS of Eq. 10 becomes singular. Therefore, we used an initial small value of $10^{-5}$ $\mu m$ for $w_P$ at time $t_0 = 10^{-6}$ s. We verified that the results are insensitive to the specific values for the IC for reasonably small chosen values of initial $w_P$ at $t_0$.

The solution of Eq. 10 provides the variation of the product width as a function of time, $w_P(t)$, and in turn, the halo width $w_H(t)$ by using Eq. 7. These are plotted in Figures 13(a) and 13(b), respectively, for two different bounds of the diffusivity parameter (corresponding to surface and bulk diffusion). These linear-scale plots show that both $w_P$ and $w_H$ increase rapidly initially, before plateauing off at longer times, as is expected for diffusion-controlled growth. Similarly, higher diffusivity values lead to greater widths at any given instant for both $w_P$ and $w_H$.

Additional insight regarding the nature of the growth is gained when these quantities are re-plotted on a log-log scale in Figures 13(c) and 13(d). The straight-line plots demonstrate a power-law relation of the kind $w = kt^n$, and we obtain the index $n$ from the slope. As indicated on these plots, $n$ seems to be inversely correlated with the parameter $D_{\mathrm{Cu}}$: $n$ ranges from 0.28 to 0.46 for $w_P$ and from 0.44 to 0.52 for $w_H$ when $D_{\mathrm{Cu}}$ is changed from $5 \times 10^{-12}$ to $5 \times 10^{-21}$ m$^2$/s, respectively. However, the dependence of $n$ on $D_{\mathrm{Cu}}$ becomes progressively weaker as larger values of $D_{\mathrm{Cu}}$ are used. We note that although the constant of proportionality $k$ depends on diffusivity in diffusion-controlled growth, the growth exponent usually does not. The unintuitive $D$-dependence of $n$ can be understood in terms of the two regimes of power-law relation between $w_P$ and $w_H$ that, as discussed earlier in connection with Figure 11(b), result from species balance considerations.

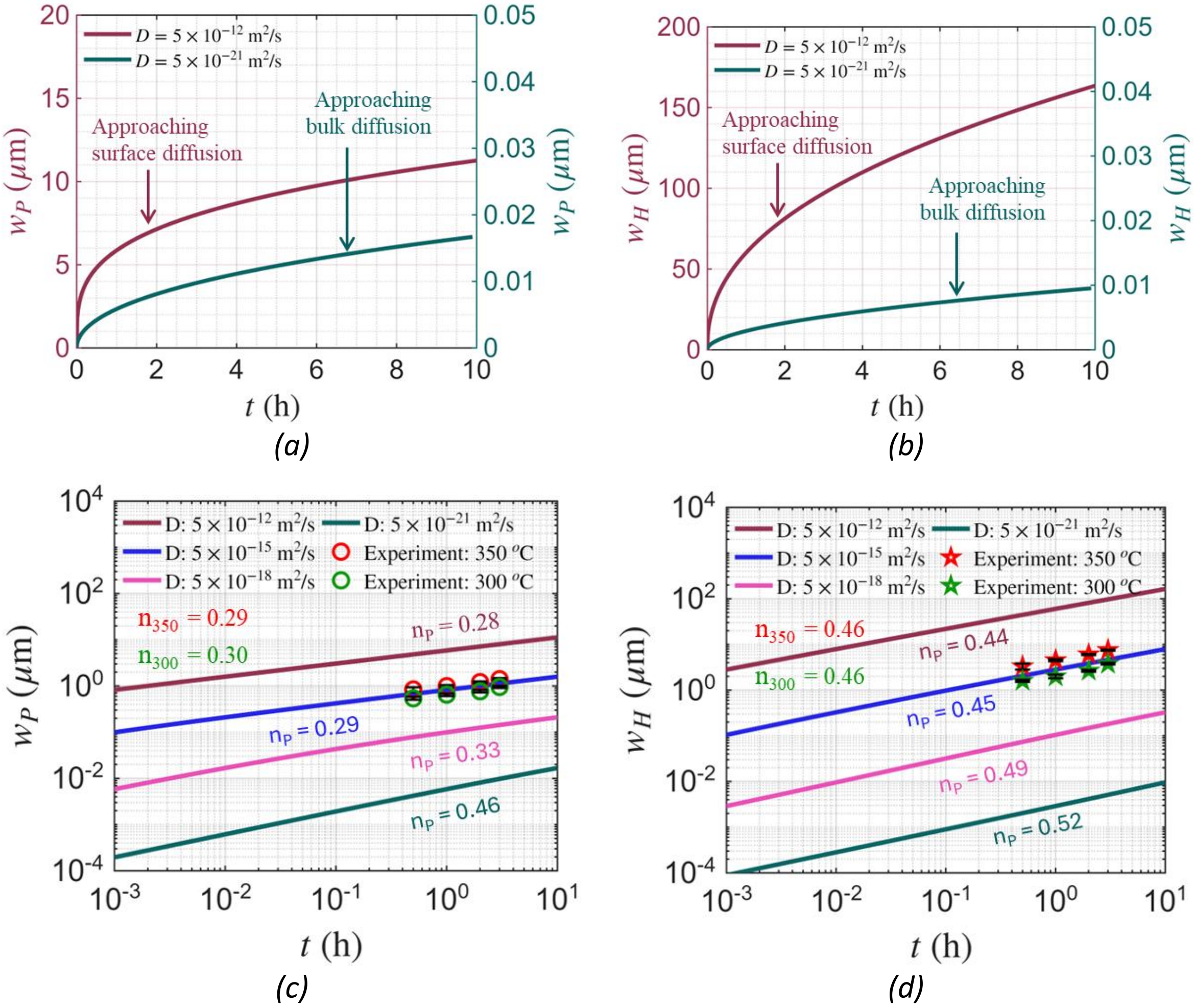

***Figure 13:*** *(a, b) Variation of the product width ($w_P$) and halo width ($w_H$) with annealing time with diffusivity as a parameter. (c, d) Corresponding log-log plot of the same. The curves are generated by numerically solving the ODE in Eq. (10); green and red points are the experimental data for 300 °C and 350 °C, respectively.*

As Figure 13(c) shows, higher $D_{\mathrm{Cu}}$ values, say, in range $10^{-15}$ to $10^{-12}$ m²/s, lead to larger product length scales, say, $\sim$100 nm (*i.e.*, $w_P/h_F > 2$) or more, and the exponent is $\sim$0.29. However, at a lower diffusivity value like $\sim 10^{-21}$ m²/s, $w_P$ is very small, $\sim$50 nm or less (*i.e.*, $w_P/h_F \sim 1$ or less), and the exponent is $\sim$0.44. Figure 13(d) shows a similar trend for the halo length scale as well. These exponent values can be understood qualitatively from the two different correlations between normalized $w_H$ and $w_P$ demonstrated in Figure 11(b). For $w_P/h_F \ll 0.2$, which also happens to be the case associated with low-$D_{\mathrm{Cu}}$ values, the second term inside the bracket in the denominator of the RHS of Eq. 10 can be dropped, and since $w_H$ scales linearly with $w_P$ in this regime, the ODE reduces to the one similar to classical Zener-type model and the growth index approaches 1/2.

For $w_P/h_F \gg 2$ associated with higher values of $D_{\mathrm{Cu}}$, $w_H$ scales as $w_P^{3/2}$ (Figure 11(b) and the leading order term in Eq. 7), which on insertion in the $w_H w_P$ term in Eq. 10, followed by integration and inversion gives rise to a $t^{2/7}$-dependence for $w_P$, which is very close to the limiting value of 0.28 obtained from Figure 13(c). Thus, the striking departure of the exponent from its classical value of 1/2 is a direct consequence of the constraints placed on growth by mass balance in a growth geometry constrained by film thickness and crystallographic aspects. These constraints create a mismatch between the dimensionality of solute transport (2d) and that of the product growth (3d) in the thin-film limit. The experimental data for both 300 °C and 350 °C also yield the exponent value close to 0.29, thereby placing them on the high-$w_P$, high-diffusivity regime.

Diffusivity is an input parameter for this model, and indeed for similar models involving diffusion-controlled growth, whose value is required for reliable predictions. Since we have experimentally measured values of $w_P(t)$ at two different temperatures and the governing ODE, the value of diffusivity can be obtained by an inverse optimization procedure. For this purpose, we used the ODE optimization tool in MATLAB which returns the best-fitting value for the $D_{\mathrm{Cu}}$ parameter by minimizing the difference between predicted and measured $w_P(t)$ values. This gives $D_{\mathrm{Cu}}$ values of 2.1×10$^{-15}$ m$^2$/s and 8.9×10$^{-15}$ m$^2$/s at 300 °C and 350 °C, respectively. These values are substantially greater than the corresponding lattice diffusivities at these temperatures ($D^L_{\mathrm{Cu}} = 3.2\times10^{-21}$ and 5×10$^{-20}$ m$^2$/s, respectively [92]). Figure 13 suggests that if bulk diffusion alone is considered, the amount of $Cu_3Si$ formed would be much less than what is observed in these experiments. Therefore, the supply of Cu must take place via high diffusivity paths like free surface, grain and interphase boundaries for the large amount of $Cu_3Si$ observed in the experiments.

GB diffusion becomes more important at lower temperatures, finer grain sizes, and, especially relevant for the present case, in immiscible systems, since GBs can accommodate much more solute than the grain interior. Niktin *et al.* [68] showed that Cu segregates strongly along the grain boundaries of Ag-rich phase in sputter deposited dilute Ag-4.9 at% Cu films. They reported that while the Ag-rich matrix contained very little Cu (∼2.2 at% Cu), depending on the nature of the GBs, the Cu concentration there varied from 15 to 75 at% Cu. These findings support the proposition that in the present case, Cu-atoms from far-field are transported to the reaction front primarily through the GB network in the Ag-rich halo. Bukaluk *et al.* [93] investigated GB diffusion of Cu through ion-plated Ag films in the 100-250 °C temperature range and reported values of $D_0 = 2.3\times10^{-9}$ m$^2$/s and $E_a = 0.68$ eV/atom for the diffusivity relation $D = D_0 \exp(-E_a/k_B T)$. The GB diffusivities calculated using these values are $2.41\times10^{-15}$ and 10$^{-15}$ m$^2$/s and $7.28\times10^{-15}$ m$^2$/s at 300 and 350 °C, respectively. The values arrived at using the approach presented above agree remarkably well with these findings.

# 4. Conclusions

We showed film-substrate reactions can be exploited to introduce local microstructural modifications in phase separating Ag-Cu thin films deposited on a Si substrate. The reaction-modified zone is made up of a central $Cu_3Si$ silicide phase surrounded by an Ag-rich matrix. This latter zone, termed as the halo, is distinct from the bulk film microstructure which consists of a phase-separated structure of Cu-rich domains embedded in an Ag-rich matrix. The influence of annealing temperature and time on the growth kinetics of these reaction-modified structures was investigated. Informed by the experimental observations, we developed a semi-analytical model that incorporates species balance in a constrained geometry and linearized flux balance. The fowling points highlight the key findings of the study:

1. Localized reaction between Cu in the film and Si-substrate near FIB milled apertures results in the formation of a central $Cu_3Si$ phase. Its electron diffraction patterns contain reflections corresponding to the fundamental η-$Cu_3Si$ hexagonal unit cell, as well as additional features like satellite spots and streaks along the $c^*$ axis.
2. Kinetics experiments show that the widths of the primary reaction product $Cu_3Si$ ($w_P$) and the halo ($w_H$) increase with increase in annealing time and temperature. At the annealing temperatures of 300 °C and 350 °C, the variation in $w_P$ and $w_H$ with time ($t$) follows the power-law relationships $w_P \propto t^{0.3}$ and $w_H \propto t^{0.46}$.
3. The $Cu_3Si$ phase grows into the Si-substrate with a characteristic V-shaped geometry where its inclined faces make an angle of $\sim$54.7° with respect to the substrate surface. This specific geometry is considered to construct a species balance equation that shows the correlation between $w_H$ and $w_P$. The general non-linear relationship between $w_H$ and $w_P$ (scaled by film thickness) is shown to have two distinct regimes of power-law relations.
4. An expression for the growth rate of $Cu_3Si$, $dw_P/dt$, is derived by modifying the classical interface species balance condition. The solute flux in the halo is determined by assuming a linear concentration gradient through a homogenized material. The resulting non-linear ODE is solved numerically to obtain $w_P(t)$ and $w_H(t)$. The calculated values of $w_P(t)$ and $w_H(t)$ fit well to a power-law relationship with time exponents of $\sim$0.29 and $\sim$0.46, respectively. For the thickness/diffusivity regime under which the experiments fall into, these exponents are very close to the asymptotic limits of 2/7and 3/7.
5. The deviation in the time exponent from the classical value of 1/2 is shown to be a consequence of the mismatch between the dimensionality of growth and that of solute transport.
6. By combining the experimental data and the model, we estimated the grain boundary diffusivity of Cu in Ag. The estimated values at 300 °C and 350 °C are $2.1\times10^{-15}$ m$^2$/s and $8.9\times10^{-15}$ m$^2$/s, respectively, which are found to be in reasonable agreement with values reported in the literature.

Although we demonstrated microstructural tuning in the Ag-Cu system, this approach can be extended to other material systems like Ag-Ni, Au-Pd, Cu-Ni, Cu-Pd (as well as ternary systems) involving at least one reactive element deposited on a Si substrate. In these systems, the reaction between the component(s) of the alloy film and the Si substrate will lead to a modification of the microstructure in the vicinity of the reaction site. Note that some of these elements are extensively used for catalysis, electronic and optical applications [94–99]. Additionally, one could choose alternative reactive substrates like Ge, Cu, GaAs which could provide further control over the reaction products and kinetics, thereby opening up newer avenues of microstructure engineering.

## CRediT authorship contribution statement

**Vivek C. Peddiraju:** Methodology, Investigation, Formal Analysis, Data Curation, Validation, Visualization, Writing – original draft; **Shourya Dutta-Gupta**: Conceptualization, Funding acquisition, Resources, Writing – review & editing; **Subhradeep Chatterjee:** Conceptualization, Resources, Formal Analysis, Supervision, Visualization, Validation, Writing – review & editing

## Declaration of competing interests

All authors declare that they have no known financial or personal relationships that could have appeared to influence the work reported in this paper.

## Data availability

All data that is relevant for drawing the main conclusions in this paper are presented in main text and supplementary materials. Any additional raw data that supports the findings of this study are available from the corresponding author upon reasonable request.

## Acknowledgements

Authors thank Prof. Saswata Bhattacharya for technical discussions. S.D.-G. would like to acknowledge the funding from SERB (Grant No. SERB/ECR/2018/002628).

Supplementary Materials

# Reaction-Diffusion Driven Patterns in Immiscible Alloy Thin Films

Vivek C. Peddiraju, Shourya Dutta-Gupta, Subhradeep Chatterjee*

Department of Materials Science and Metallurgical Engineering,

Indian Institute of Technology Hyderabad, Kandi, Sangareddy, Telangana 502285, India

Email: subhradeep@msme.iith.ac.in

This document contains additional text and figures which support the data provided in the main manuscript.

**Image autocorrelation function (ACF) analysis and linear chord length distribution (CLD) analysis:**

The ACF analysis, also termed as the two-point statistics, determines the probability that two randomly selected points at a given distance to belong to the same phase. The CLD on the other hand is based on the conventional linear intercept method which returns the distribution of chord lengths (intercepts) that lie within a phase of interest. The reason for utilizing two different metrics is that the ACF analysis fails to detect the presence of multiple significant length scales (*e.g.*, a bimodal distribution), as well as any morphological anisotropy present within the microstructure. The CLD reveals such features readily as it is obtained along different orientations within the microstructure. Note that in general, there is a wide variation in the distribution of chord lengths for even a simple geometric shape like circle. However, the mean value always scales with the actual size of the feature and it only should be considered as a meaningful representation of length scale.

Both the analysis methods are implemented in MATLAB. First, the SEM/TEM images are smoothened using a gaussian filter. This operation smears out random noises present within the captured images so that constituent domains (say, Ag-rich or Cu-rich) within the image can be demarcated unambiguously. The gray scale image is then converted to binary images via a thresholding operation for subsequent analysis. The volume fractions of the two constituent phases are obtained directly from the binarized micrograph by calculating the fraction of black or white pixels within a region of interest.

The ACF of an image in vectoral representation is [1–3]:

$$ACF(\boldsymbol{r}) = \frac{1}{N}\sum_{\boldsymbol{r}}\sum_{\boldsymbol{r'}} I(\boldsymbol{r'})I(\boldsymbol{r'}+\boldsymbol{r}) - \langle I^2 \rangle \qquad (S1)$$

where $\boldsymbol{r}$ is a position vector pointing to pixel $(x, y)$ with respect to a chosen origin, $\boldsymbol{r'}$ is displacement vector from $\boldsymbol{r}$, $N$ is total number of pixels in image and $I$ is the intensity value at any position within the image. The ACF matrix can be computed directly by the double summation of image intensity as per the following expression:

$$ACF(x,y) = \sum_{i=1}^{px}\sum_{j=1}^{py} I(x,y)\, I(x+i, y+j) - \langle I^2 \rangle \qquad (S2)$$

where $px$ and $py$ represent the total number of pixels along each dimension in the original image for which we are computing autocorrelation matrix. The ACF of an image can also be calculated more efficiently by reciprocal space methods based on the Wiener-Khinchin theorem [2] which states that the ACF and the image power spectrum together form a Fourier transform pair. Thus, the image ACF is computed as:

$$ACF(x,y) = \boldsymbol{\mathcal{F}}^{-\mathbf{1}}\big(\boldsymbol{\mathcal{F}}(\boldsymbol{I}).\boldsymbol{\mathcal{F}}^{*}(\boldsymbol{I})\big) \qquad (S3)$$

where $\boldsymbol{\mathcal{F}}$ and $\boldsymbol{\mathcal{F}}^{*}$ represent the discrete Fourier transform operation and its conjugate, and $\boldsymbol{\mathcal{F}}^{-\mathbf{1}}$ stands for the inverse transform. Figure S1 shows the equivalence of the two methods of computing the ACF (Eq. S2 and Eq. S3) using a simulated microstructure of interconnected domains. Circular averaging of the ACF is carried out by dividing the region of interest into multiple annular segments of constant width at discrete radial distances from the origin. The circularly averaged intensity profile of the computed correlation matrix provides the necessary quantitative data about microstructure. The first zero of the circularly averaged ACF profile corresponds to characteristic length scale of two-phase microstructure. Figures S1(b-d) establish the equivalence of these two methods.

In the CLD analysis, the length scale is computed from the line profiles, drawn on the binary micrograph along 0°, 45°, 90° and 135° orientations. The Cu domain size, considering that it is black in binary image, is computed by counting the number of pixels in between the event where intensity switches from 1 to 0 and 0 to 1 successively. This is accomplished by using the *strfind* tool in MATLAB.

Accuracy and equivalence of these two methods (ACF and CLD) is further verified by tracking the time dependence of the length scale in a system undergoing coarsening after initial spinodal decomposition. Simulated microstructures and the corresponding time dependence of the length scale are presented in Figure S2.

## Supporting Figures

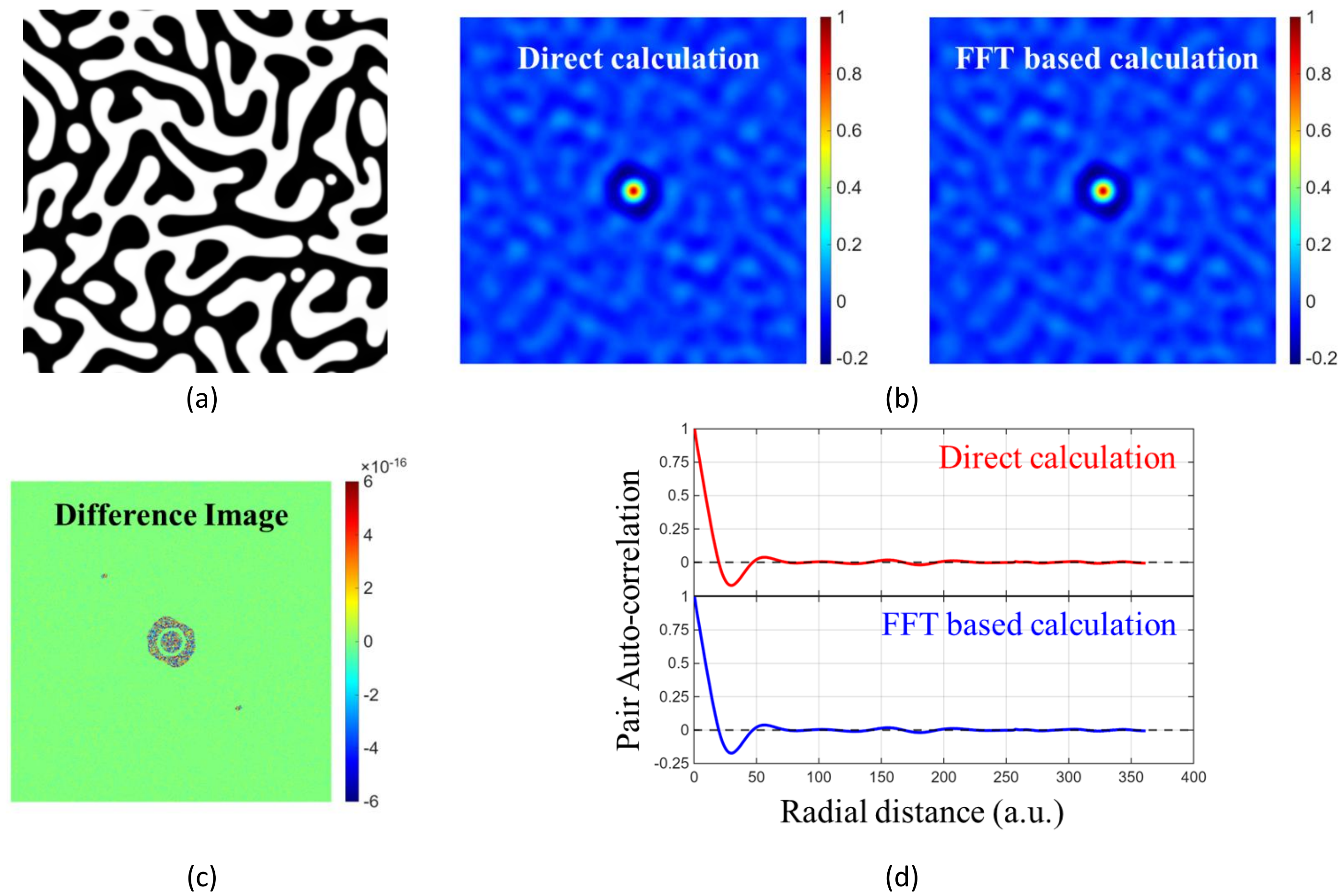

***Figure S1:*** *Equivalence of two different methods to compute the ACF (a) Test image (simulated) for which ACF is determined. Black and white pixels represent the two phases. (b) ACF calculated by the direct (real-space) method (Eq. S2) the Fast Fourier Transform (FFT) based technique (Eq. S3). (c) Difference map between the methods. (d) Circularly averaged radial ACF profiles for both methods (shown separately to avoid overlap).*

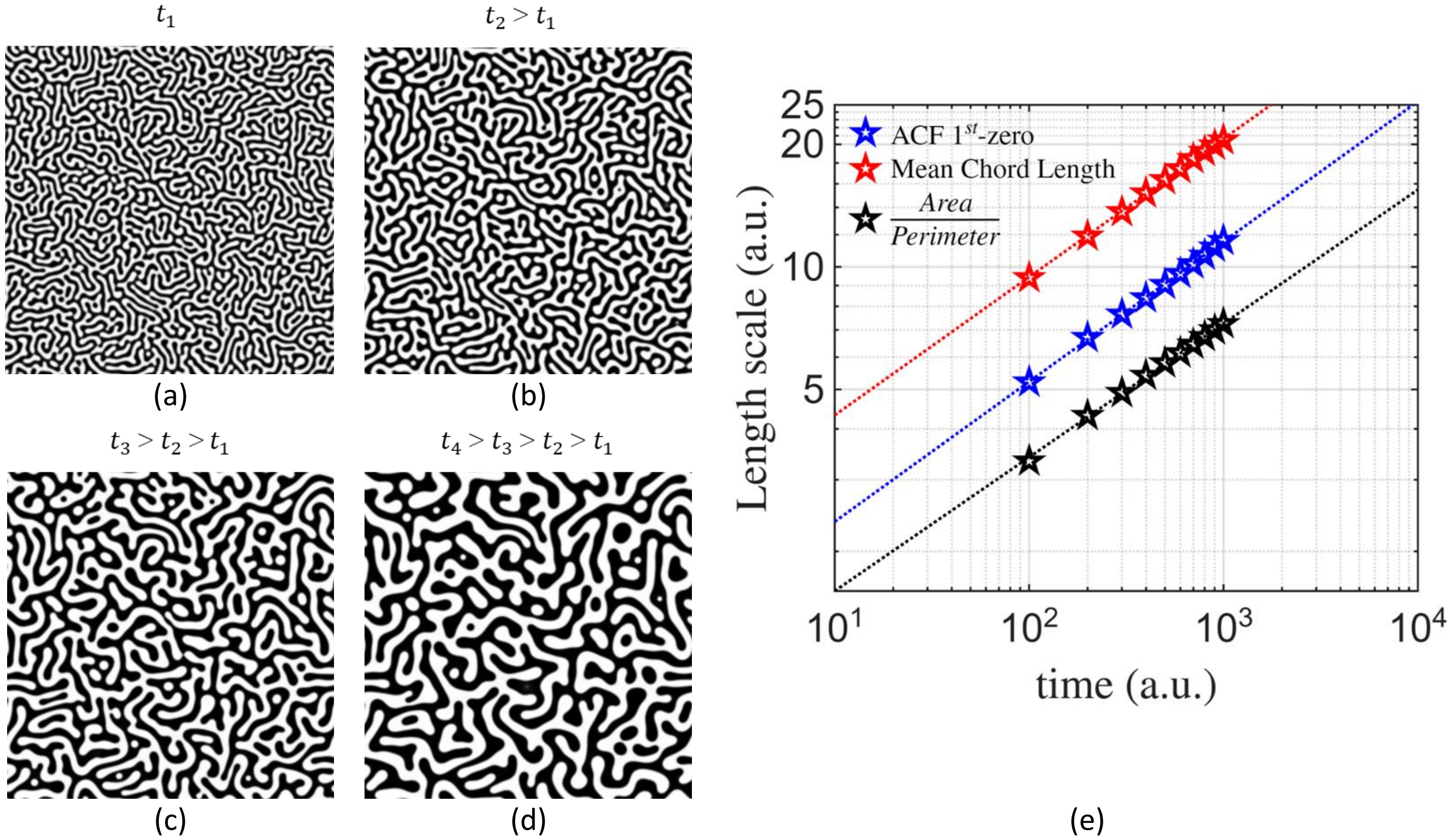

***Figure S2:*** *(a-d) Simulated test images with progressively coarser microstructures. (e) Variation of the characteristic length scale ($\lambda$) as a function of time step ($t$) on a log-log scale using both ACF and CLD methods. For both, the slope is close to $1/3$, which is the expected value of the time exponent $n$ for bulk diffusion controlled LSW-like coarsening process. $S_V^{-1}$ is the ratio of perimeter to area of one of the phases (e.g., the bright one).*

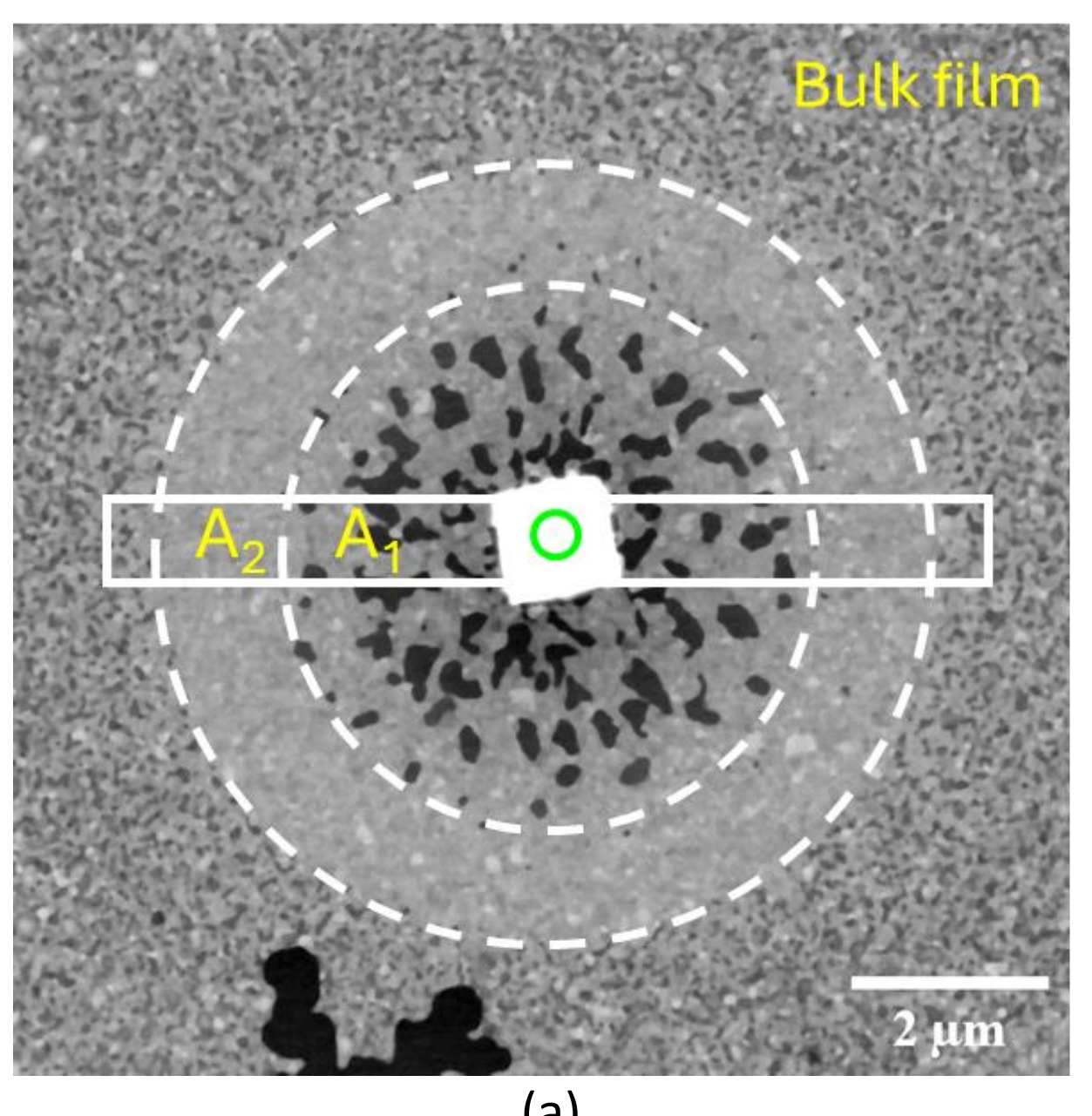

(a)

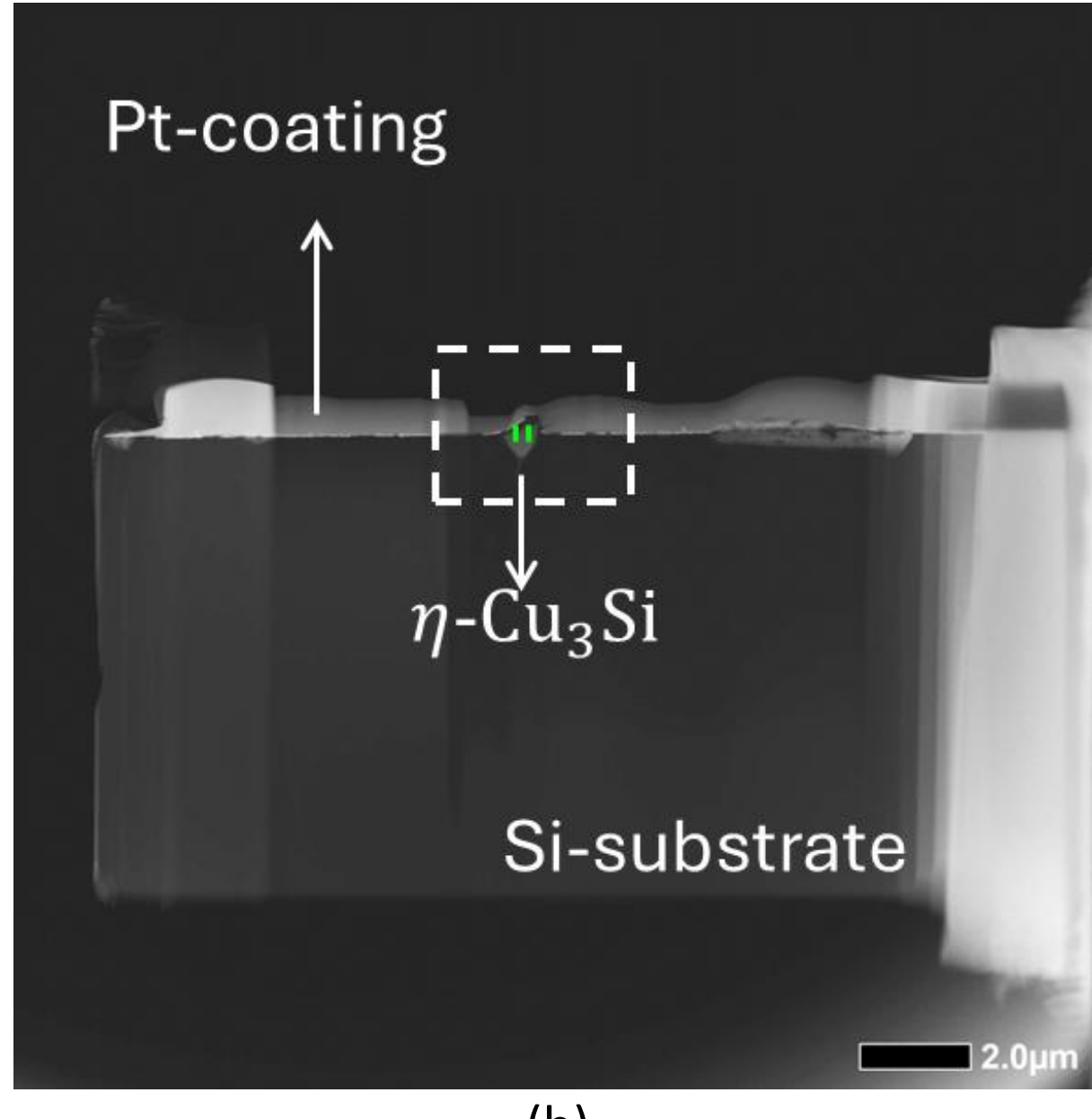

(b)

***Figure S3:*** *(a) Location and geometry of the FIB lamella lift-out (rectangle bound by solid white lines) on a representative SEM-BSE micrograph with a central $Cu_3Si$ particle (bright) surrounded by the halo structure. (b) A low magnification STEM image of the specimen extracted using $Ga^+$ FIB milling. $Cu_3Si$ formed near the aperture by film-substrate reaction is indicated by the dashed white box. A magnified view is shown in Figure 2(a) of the main manuscript. Green circle in (a) and vertical lines in (b) indicate the original location of the aperture which is completely filled by the central silicide particle.*

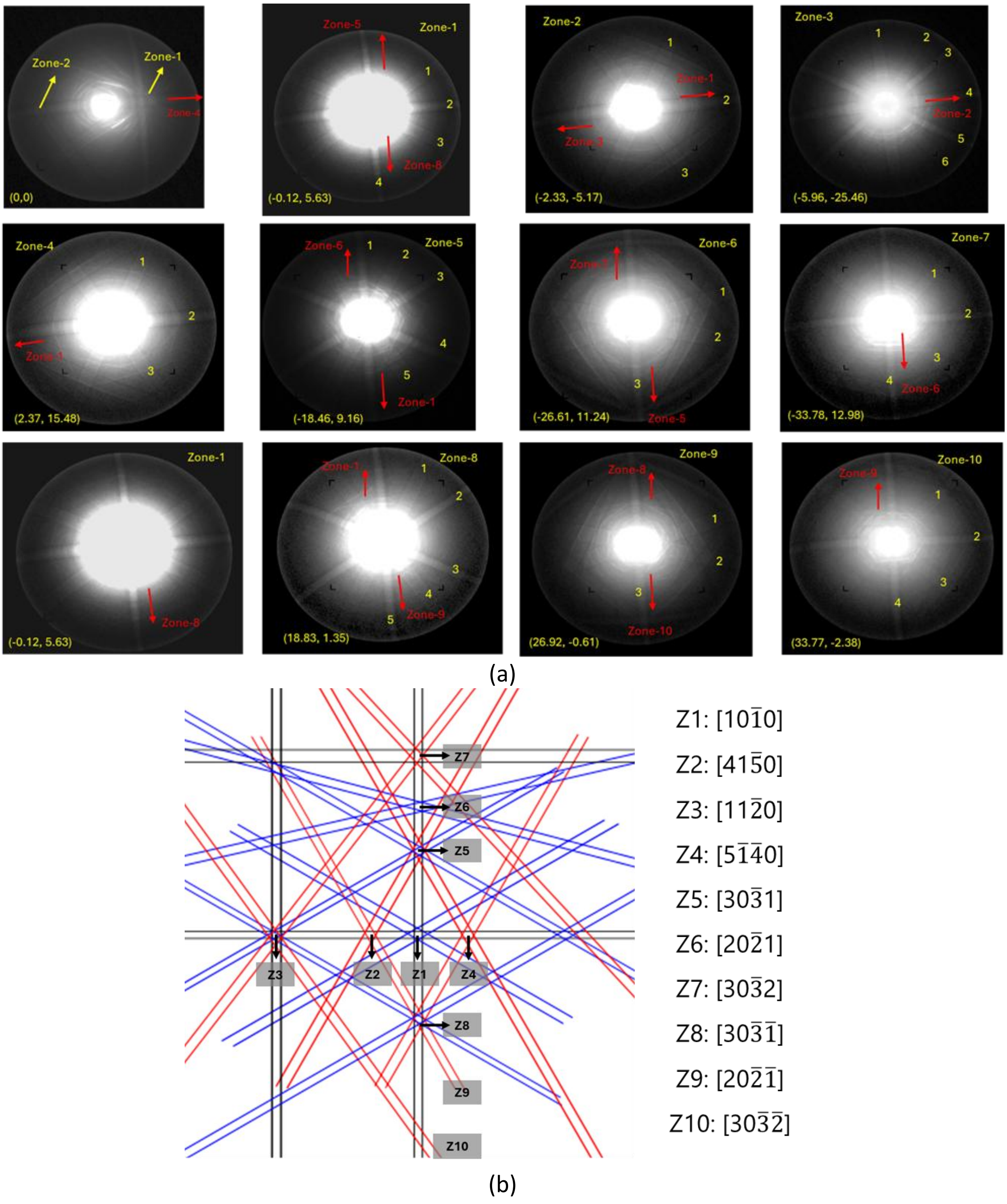

**Figure S4:** (a) A collection of Kikuchi bands corresponding to various zones obtained by systematic tilting the $\eta$-$Cu_3Si$ particle. Zone centers are shown in yellow and zones approached are in red. The $x$- and $y$-tilt values are mentioned as $(T_x, T_y)$ pairs on each image. (b) Schematic view of the larger Kikuchi space obtained by combining information from these images.

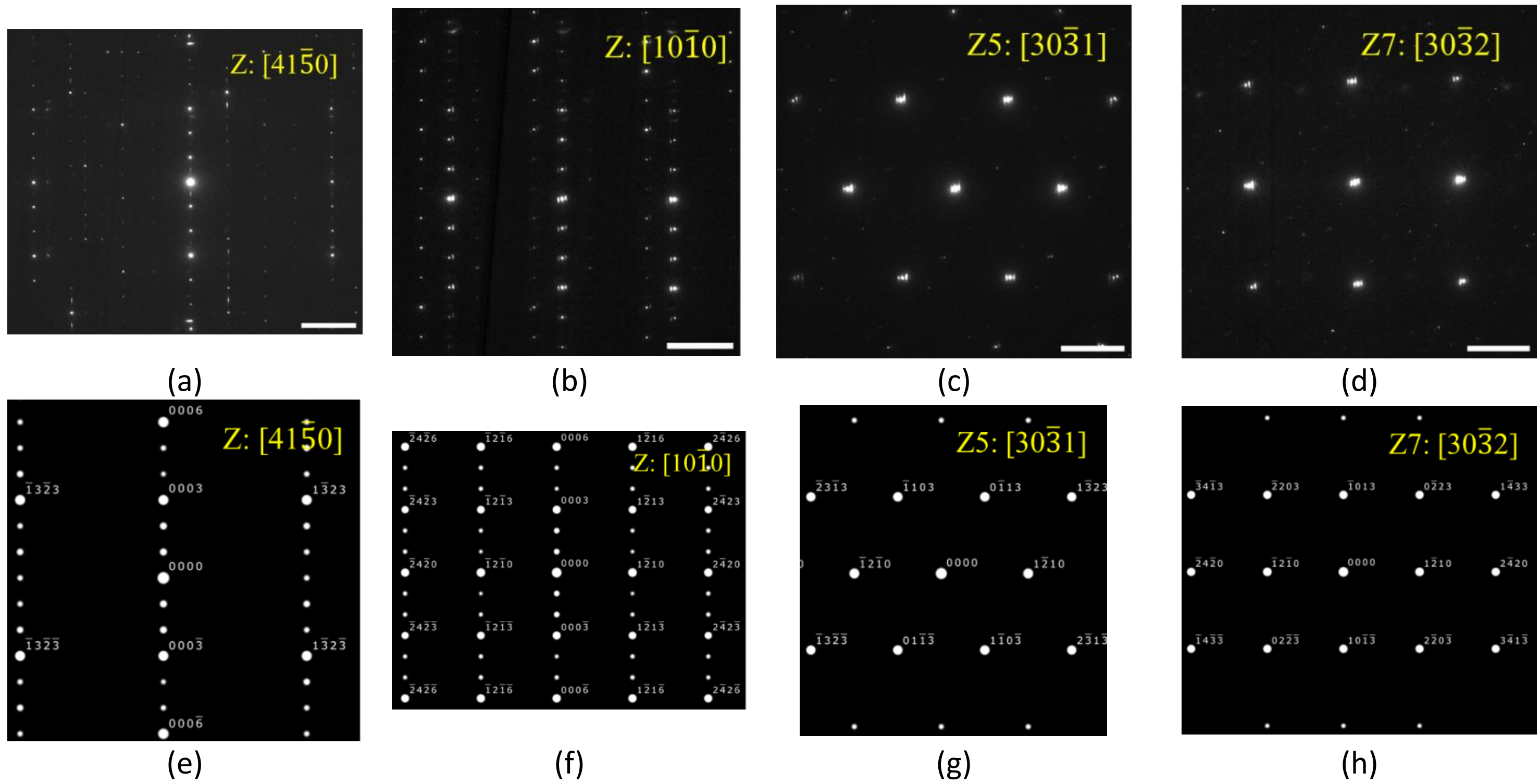

***Figure S5:*** *(a–d) Experimentally recorded and (e–h) simulated electron diffraction patterns obtained from the $Cu_3Si$ particle. The simulated patterns are generated using the SingleCrystal® 5 program in the CrystalMaker® software suite. As shown in Figure 2 of the main manuscript, the high temperature η-$Cu_3Si$ structure is used for generating these patterns. Scale bar corresponds to 3 1/nm.*

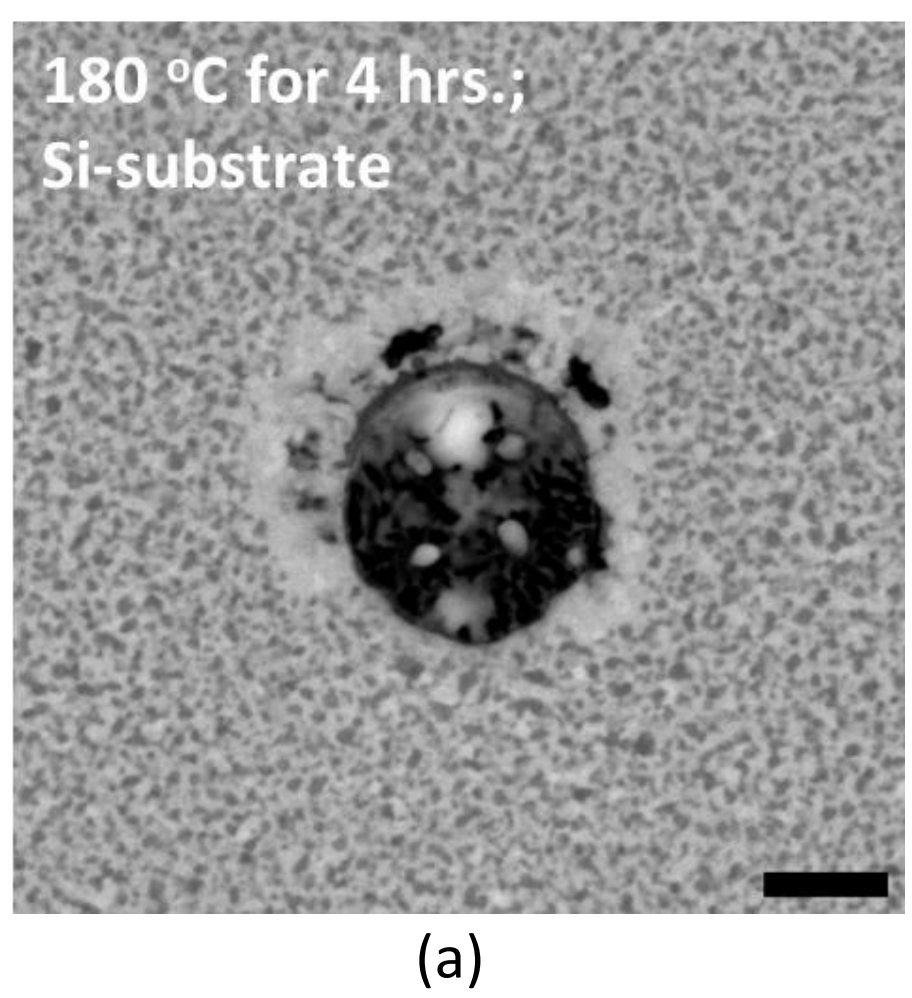

(a)

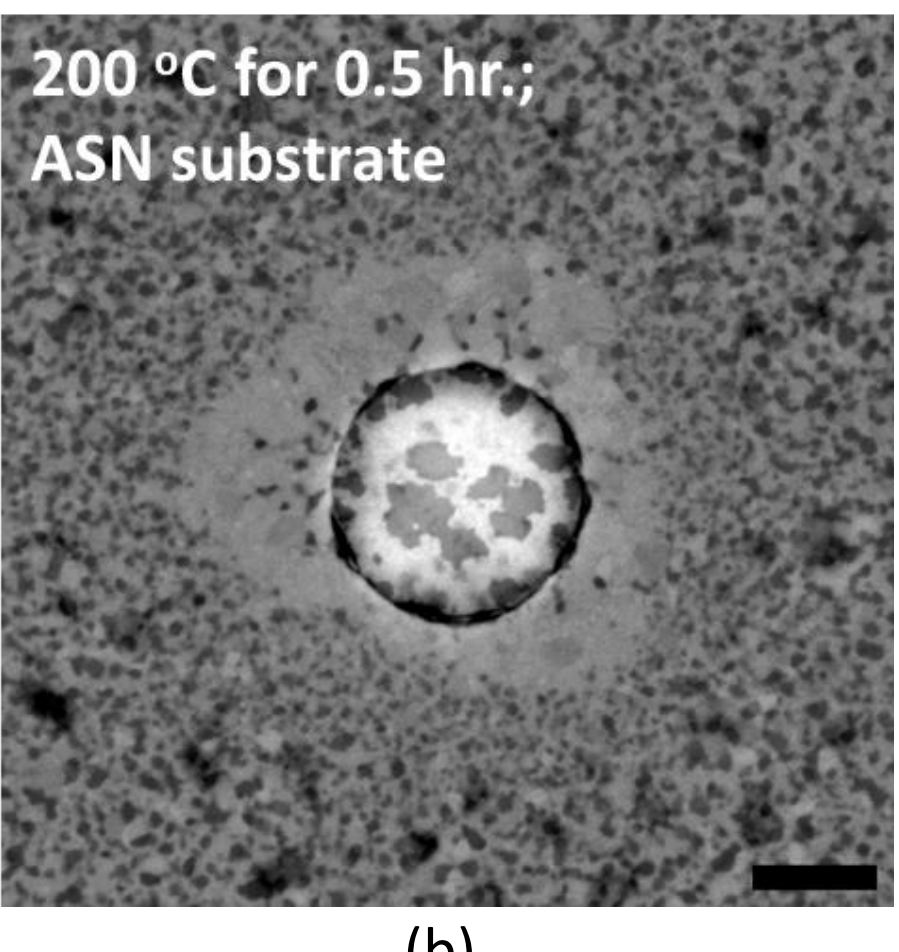

(b)

***Figure S6:*** *(a) Microstructure near the aperture after annealing at 180 °C for 4 h. Onset of local reactions is evident from irregular particles formed along the inner rim of the aperture and the Ag-rich layer adjacent to the rim. Phase separation in the bulk film is consistent with the GIXRD results of Figure 3 of the main manuscript. (b) Faster reaction and growth kinetics at 200 °C lead to a greater amount of silicide along with a more continuous halo in a shorter duration (0.5 h). All the micrographs are captured in SEM-BSE mode. Scale bar corresponds to 500 nm.*

| **Annealing conditions** | 0.5 hour | 1 hour | 2 hours | 3 hours |
|---|---|---|---|---|
| 250 $^{o}$C | (e) | (f) | (g) | (h) |
| 300 $^{o}$C | (i) | (j) | (k) | (l) |

***Figure S7:*** *Bulk film microstrucutres showing the formation of Ag-rich (bright contrast) and Cu-rich(dark contrast) domains during annealing of as-depostied metastable sinlge phase Ag-Cu films. All images are captured in SEM-BSE mode. Scale bar corresponds to 250 nm.*

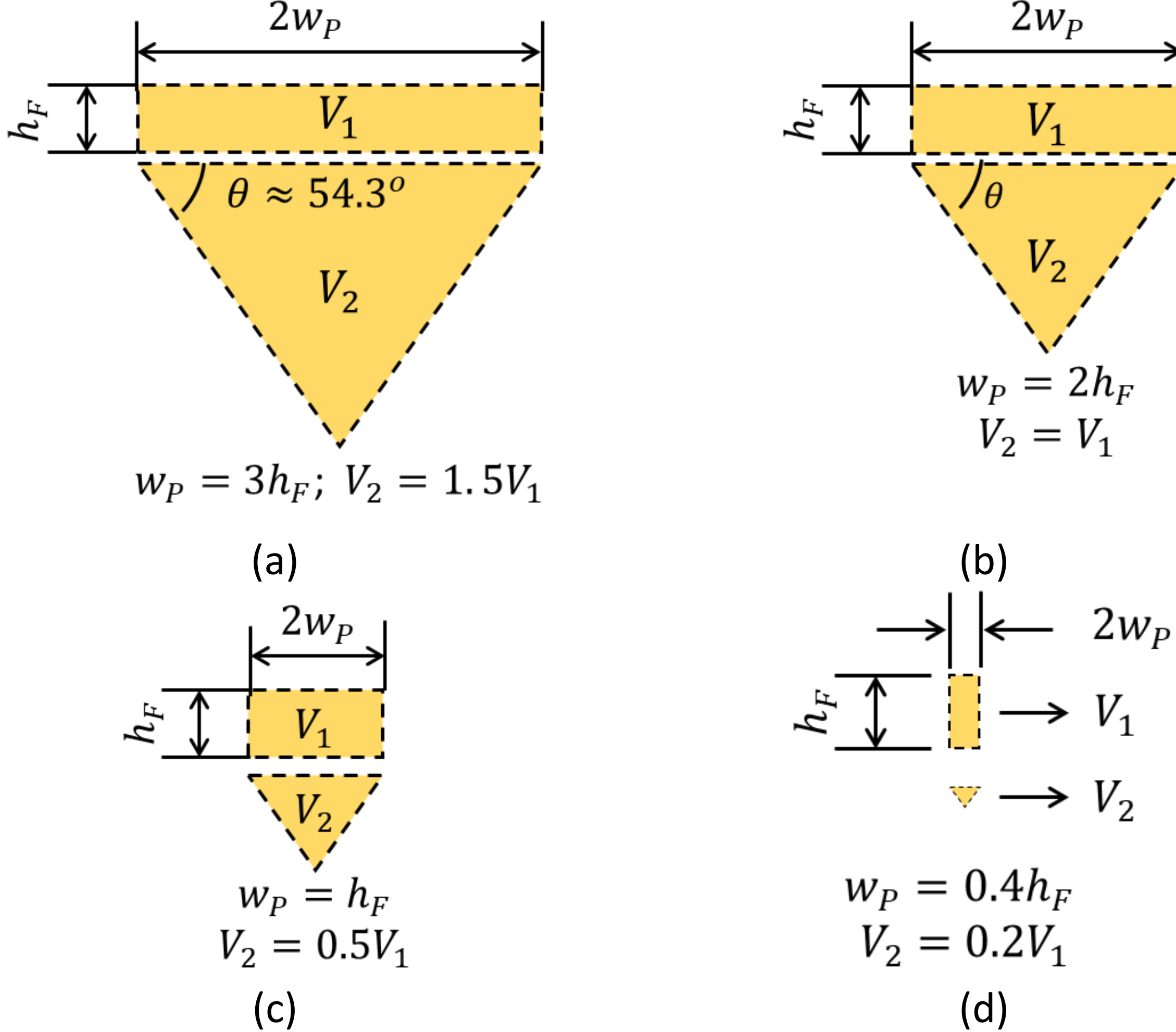

***Figure S8:*** *Schematic illustration of the product volume into two parts with volume $V_1$ and $V_2$. The top part has a disc or cylindrical geometry with a thickness of $h_F$ and radius of $w_P$: $V_1 = \pi w_P^2 h_F$.The bottom part grows into the substrate and idealized as an inverted cone with base radius $w_P$ and height $w_P \tan\theta$, giving volume $V_2 = \frac{1}{3}\pi \tan\theta\, w_P^3$. Thus, $\frac{V_2}{V_1} = \frac{w_P}{2h_F}$. The sketches in (a-d) illustrate the variation of the product geometry for different values of this ratio. During the initial stages and/or for thicker films, $w_P \ll h_F$ and the $V_2$-contribution is much smaller, reducing the growth approximately to a 1-d problem where $w_H$ scales linearly with $w_P$. At larger widths, the $V_2$ contribution becomes larger, resulting in non-linear relationship between $w_H$ and $w_P$. For clarity, $V_1$ and $V_2$ are shown separated by a gap here.*